\documentclass[11pt,letterpaper,leqno]{article}

\usepackage{amsmath}
\usepackage{epsfig,graphics}
\usepackage{etoolbox}
\usepackage{palatino}
\usepackage{subcaption}
\usepackage{multicol}
\usepackage[hyperfootnotes=false]{hyperref} 
\hypersetup{pdfborder=0 0 0} 
\hypersetup{colorlinks=true, urlcolor=MidnightBlue, citecolor=MidnightBlue, linkcolor=MidnightBlue}
\usepackage{mathrsfs}
\usepackage{multirow}
\usepackage{graphicx}
\usepackage{caption}
\usepackage{subcaption}
\usepackage[bottom,multiple]{footmisc}
\usepackage{ragged2e}

\usepackage[usenames,dvipsnames,svgnames,table]{xcolor}

\makeatletter
\def\@footnotecolor{gray!70!black}
\define@key{Hyp}{footnotecolor}{%
 \HyColor@HyperrefColor{#1}\@footnotecolor%
}
\patchcmd{\@footnotemark}{\hyper@linkstart{link}}{\hyper@linkstart{footnote}}{}{}
\makeatother
\hypersetup{footnotecolor=black}

\makeatletter
\renewcommand\@biblabel[1]{}

\makeatother

\usepackage{amsthm, thmtools}
\usepackage{
nameref,
hyperref,
cleveref,
}

\usepackage{amscd}
\usepackage{amsthm}
\usepackage{amsfonts}
\usepackage{tikz}
\usepackage{float}
\usetikzlibrary{calc,intersections,through,backgrounds,shadings}
\usetikzlibrary{shapes.geometric}
\usetikzlibrary{through}
\usetikzlibrary{arrows.meta}
\usepackage{multirow}
\usepackage{mathtools}
\usepackage{amssymb}
\usepackage{caption}
\usepackage{xfrac}  
\usepackage[margin=.9in]{geometry}

\newcommand\nc{\newcommand}
\nc\on{\operatorname}

\theoremstyle{plain}

\newtheorem{prop}{Proposition}

\usepackage{pgfplots}

\theoremstyle{definition}

\usepackage{epigraph}
\usepackage{natbib}
\usepackage{graphicx}
\usepackage{placeins}
\graphicspath{ {images/} }
\usepackage{etoolbox}
\usepackage{amsmath}

\usepackage[utf8]{inputenc}
\usepackage[english]{babel}
 \usepackage{bm}
\usetikzlibrary{shapes.geometric}
\newtheorem*{mydef}{Definition (Competitive Equilibrium)}

\newcommand{\Zeta}{\bar{z}_{\mathrm{AI}}}
\newcommand{\zbar}{z_{\mathrm{AI}}}

{ }

\newcommand{\post}{*}

\usepackage{enumitem}
\usepackage{chngcntr}
\usepackage{apptools}
\AtAppendix{\counterwithin{prop}{section}}
\AtAppendix{\counterwithin{lem}{section}}
\AtAppendix{\counterwithin{cor}{section}}
\AtAppendix{\counterwithin{cl}{section}}
\usepackage{booktabs}

\begin{document}
\title{Artificial Intelligence in the Knowledge Economy\thanks{IESE Business School (eide@iese.edu and etalamas@iese.edu). We extend our deepest gratitude to Luis Garicano, whose invaluable suggestions have fundamentally shaped the presentation of this work. We have also benefited enormously from the comments of Ricardo Alonso, Nano Barahona, Markus Brunnermeier, Nicol\'{a}s Figueroa, Andrea Galeotti, Ben Golub, Jason Hartline, Frank Levy, Niko Matouschek, Mike Ostrovsky, Crist\'{o}bal Otero, Sebasti\'{a}n Otero, Esteban Rossi-Hansberg, four anonymous referees, and seminar participants at the various venues where we presented this work. We acknowledge the financial support of IESE through the High Impact Initiative-course 2024/2025. We declare we have no relevant or material financial interests that relate to the research described in this paper. Edited by Esteban Rossi-Hansberg.}}
  
\author{Enrique Ide \& Eduard Talam\`{a}s}
\date{\today}

\maketitle 

\thispagestyle{plain}

\begin{abstract} 
Artificial Intelligence (AI) can transform the knowledge economy by automating non-codifiable work. To analyze this transformation, we incorporate AI into an economy where humans form hierarchical organizations: Less knowledgeable individuals become “workers” doing routine work, while others become “solvers” handling exceptions. We model AI as a technology that converts computational resources into “AI agents” that operate autonomously (as co-workers and solvers/co-pilots) or non-autonomously (solely as co-pilots). Autonomous AI primarily benefits the most knowledgeable individuals; non-autonomous AI benefits the least knowledgeable. However, output is higher with autonomous AI. These findings reconcile contradictory empirical evidence and reveal tradeoffs when regulating AI autonomy. \end{abstract} 

 \newpage

\section{Introduction} 

Recent developments in Artificial Intelligence (AI) are driving a new wave of automation, enabling machines to perform sophisticated knowledge work such as coding, research, and problem-solving. While the potential of this technology to reshape the landscape of work is undeniable, its precise implications remain the subject of growing controversy \citep{brynjolfsson,johnson2023power,acemoglu2024simple,autor2024applying,korinek}. This controversy arises from two factors. First, it is unclear whether lessons from previous waves of automation can inform our understanding of AI's effects \citep{muro,agrawalintro}. Second, because AI is still in its infancy and individuals and firms are experimenting with its use \citep{mc2023AI,humlum2024adoption}, current empirical evidence cannot yet fully capture its equilibrium effects. 
 
In this paper, we develop a framework for studying the equilibrium effects of AI on the knowledge economy.\footnote{According to data from the U.S. Bureau of Labor Statistics, as of December 2024, over 100 million people in the U.S. were employed in cognitive or knowledge-based roles \citep{fred}.} Our approach is grounded in two insights. First, the defining characteristic of knowledge work is that production know-how is predominantly tacit (or non-codifiable)---developed through repeated observations of practical successes and failures---and therefore inherently embodied in individuals \citep{polanyi,garicano2000hierarchies,garicanowu2012}. As a result, individuals’ time and knowledge are critical bottlenecks, making firms central to organizing production \citep{garicano2015knowledge}. Second, unlike traditional automation technologies, AI learns by example, uncovering patterns that cannot be codified into explicit rules \citep{mitchell,autor2024applying,brynjolfsson2023generative}. This ability allows AI systems to acquire tacit knowledge and perform cognitive, non-codifiable tasks (see Table \ref{table:taxonomy}). By automating such work, AI has the potential to alleviate knowledge bottlenecks, opening the door to a fundamental reorganization of production.

\begin{table}[!b]
\centering
\renewcommand{\arraystretch}{1.1} 
\begin{tabular}{c|cc}
\toprule
            & \textbf{Codifiable}        & \textbf{Non-Codifiable} \\ \hline
\textbf{Manual}      & Robots            & Physical AI   \\
\textbf{Cognitive}   & Enterprise Software         & AI             \\ \bottomrule
\end{tabular}
  \captionsetup{justification=centering}
\caption{The Distinguishing Feature of AI \\  \justifying 
 \vspace{0.5mm}
\footnotesize \noindent \textit{Notes.} The defining feature of AI is its ability to perform cognitive work that relies on non-codifiable knowledge. This distinguishes it from enterprise software and robots, which automate codifiable cognitive and manual work, respectively. AI also differs from "Physical AI" \citep{nvidia2}, a concept that envisions robots powered by AI, which may, in the future, be capable of automating manual work that requires non-codifiable knowledge, such as caregiving or construction.} \label{table:taxonomy}
\end{table}

More precisely, we incorporate AI into a canonical model of the knowledge economy: The knowledge hierarchies first introduced by \cite{garicano2000hierarchies}. This model emphasizes the central role of tacit knowledge in shaping organizations and labor market outcomes. Within this framework, we study the equilibrium effects of AI on occupational choice, organizational structure, and the distribution of labor income. Our analysis highlights that introducing AI agents is fundamentally different from expanding the human labor force, and that AI's impact crucially depends on its problem-solving capabilities and its degree of autonomy. 

Our starting point---the pre-AI economy---is the baseline model of \cite{garicano2004inequality}, \cite{antras2006offshoring}, and \cite{fuchs2015optimal}. Labor and knowledge are the sole inputs in production. Humans are endowed with one unit of time and are heterogeneous in terms of knowledge. Individuals use their time to pursue production opportunities that require cognitive work---such as handling customer support, drafting contracts, or designing products---but encounter problems of varying difficulty during the production process. Output is produced when an individual successfully solves the problem she confronts, which occurs when her knowledge exceeds the problem's difficulty. If a human cannot solve a problem on her own, she may seek help from another human. Help, however, is costly in terms of time. Knowledge is tacit because matching problems to those who can solve them is difficult, and experts cannot outline a plan of action in advance. In this sense, individuals ``know more than they can tell'' \citep{polanyi}.

In the competitive equilibrium of the pre-AI economy, humans either pursue production opportunities on their own or join hierarchical firms to optimize the use of their time and knowledge. These firms exhibit \textit{management by exception}: Less knowledgeable individuals become ``workers'' who pursue production opportunities, while more knowledgeable individuals become ``solvers'' specializing in assisting workers with the exceptional problems they are unable to solve. This structure effectively shields the most knowledgeable individuals from routine work, enabling them to focus on their comparative advantage: solving complex problems.\footnote{There is both anecdotal and systematic empirical evidence showing the emergence of such ``knowledge hierarchies'' \citep[see, e.g.,][]{garicanohubbard,caliendo0,caliendo1,caliendo2}. For instance, \cite*{sloan}, a former head of General Motors (GM), once wrote: ``We do not do much routine work with details. They never get up to us. I work fairly hard, but on exceptions.”} 

Our innovation is to incorporate AI into this otherwise canonical setting. We base our model on three key developments in modern AI, documented in Section \ref{sec:motivation}. First, unlike earlier automation technologies, AI—like humans—can acquire tacit knowledge. However, unlike humans---who are constrained by their individual time---firms can leverage computational resources (“compute”) to use AI's knowledge at scale. Second, the rise of flexible, general-purpose foundation models—adaptable to diverse tasks with minimal additional training—enables firms to rely on a shared AI model. Third, technology firms are developing and starting to deploy AI agents that can do cognitive work autonomously. These agents do not just passively answer questions but can pursue projects independently.

Guided by these developments, we model AI as a technology available to all firms that converts compute into AI agents, all endowed with the same exogenously fixed level of knowledge. As a benchmark, we first consider the case where these agents can do exactly the same as humans. These AI agents are “autonomous” because they can function both as “co-workers” (pursuing production opportunities) and as “solvers/co-pilots” (providing advice). A key contribution of our analysis is to contrast this benchmark with the case of “non-autonomous” AI, where—for technological or regulatory reasons—AI agents are restricted to acting solely as co-pilots.

In the post-AI economy, firms decide their organizational structure and whether to integrate AI agents into their production processes. The total amount of compute available in the economy is exogenous, while its price is endogenously determined in equilibrium. We assume that compute is abundant relative to human time, making humans the binding constraint in human-AI interactions. This assumption reflects the exponential growth in computational capacity over the past two centuries \citep{nordhaus} and the fact that current AI models process information and generate outputs 10 to 100 times faster than humans \citep{amodei}.

Regarding autonomous AI, we show that its impact critically depends on whether it is ``basic'' or ``advanced.'' The introduction of basic autonomous AI---which creates AI agents with the knowledge of pre-AI workers---induces the most knowledgeable pre-AI solvers to focus on supporting AI-driven production. This reallocation forces the least knowledgeable human workers to seek assistance from less knowledgeable individuals, pushing the marginal pre-AI workers to take specialized problem-solving roles. In other words, basic autonomous AI displaces humans from routine work towards complex problem-solving. In this scenario, the least knowledgeable individuals are harmed by the technology---they face competition from AI in production work and must rely on less knowledgeable individuals for support---whereas the most knowledgeable individuals benefit by using AI to leverage their expertise at a low cost.

In contrast, the introduction of an advanced autonomous AI---which creates AI agents with the knowledge of pre-AI solvers---induces the least knowledgeable humans to seek assistance from AI instead of humans. Consequently, the marginal pre-AI solvers are pushed into routine work, displacing humans from complex problem-solving. In this scenario, both the least and the most knowledgeable individuals benefit from the technology. The least knowledgeable gain access to relatively cheap AI agents that can help them solve complex problems, while the most knowledgeable continue to thrive by leveraging advanced AI agents for routine work.

Recent anecdotal evidence from professional services industries illustrates these reorganizations. For example, \cite{beane2024inverted} document that junior analysts in investment banking are increasingly distanced from senior partners as AI takes over tasks that once required junior involvement. Similarly, a 2022 survey of M\&A lawyers found that AI, by automating routine knowledge work such as document review, has allowed junior lawyers to shift their focus to more complex responsibilities, including client engagement and legal analysis \citep{litera}.

The impact of AI on labor outcomes also critically depends on its autonomy. We show that non-autonomous AI tends to benefit the least knowledgeable individuals the most, in contrast to autonomous AI, which mainly benefits the most knowledgeable. However, overall output is higher with autonomous AI than with non-autonomous AI. 

Intuitively, non-autonomous AI primarily benefits the least knowledgeable because it does not compete with them for production work and provides a means of solving complex problems without human assistance. Its lower opportunity cost relative to autonomous AI—stemming from its inability to perform production work independently—reinforces this advantage, allowing the least knowledgeable to capture a larger share of AI-generated benefits. For the most knowledgeable individuals, the effect is reversed. Non-autonomous AI is less advantageous than autonomous AI because it lacks the capacity to handle routine work and competes with them for advisory roles.

Our findings reveal a trade-off between output and inequality when regulating AI autonomy. They also reconcile seemingly contradictory empirical findings on AI’s impact on work. On the one hand, some evidence suggests that AI disproportionately benefits the least knowledgeable individuals and reduces performance inequality \citep[e.g.,][]{dell2023navigating,noy2023experimental,peng2023impact,wiles2023algorithmic,demming,brynjolfsson2023generative}. On the other hand, there is also evidence that AI complements high-skilled knowledge workers while substituting for their low-skilled counterparts \citep{berger2,azar2024AI}. Although these findings may seem contradictory, they can be rationalized by our framework as we explain in Section \ref{sec:conclusion}: They align with the distinct impacts of non-autonomous AI and basic autonomous AI, respectively.

\subsection*{Related Literature}

This paper builds on the literature on knowledge hierarchies—which investigates how tacit (or non-codifiable) knowledge shapes organizational structures and labor market outcomes—to deepen our understanding of the effects of automation. Specifically, we explore the implications of AI, a new technology that can automate work traditionally reliant on tacit knowledge.

The literature on knowledge hierarchies originates with \cite{garicano2000hierarchies}, who introduces the model and shows that knowledge hierarchies are optimal when production requires the application of tacit knowledge.  \cite{garicano2004inequality,garicano2006organization} extend this model to include heterogeneous agents, exploring how the endogenous organization of knowledge work affects labor income inequality. \cite{fuchs2015optimal} further develop this literature by characterizing the equilibrium contractual arrangements when there is asymmetric information about knowledge.\footnote{Other important contributions to this literature include \citet{garicanohubbard2,garicano2009specialization}, \cite{garicano2012growth}, \cite{bloom2014distinct}, \cite{garicano2018earnings}, \cite{caicedo2019learning}, \cite{gumpert2022firm}, \cite{tamayo2024organizational}, and \cite{carmona2024improving}.} Building on this foundation, we develop a framework to examine how modern AI can transform knowledge work.

In the context of knowledge hierarchies, the most closely related paper is \cite{antras2006offshoring}. They study the effects of offshoring by comparing the equilibrium of a closed economy with one in which firms can form international teams. Our paper differs from theirs in two key respects. First, while offshoring gives firms access to a human population with varying knowledge levels, AI allows firms to automate knowledge work at scale. As discussed in Section \ref{sec:conclusion}, this leads to qualitatively different outcomes.  Second, we introduce AI-specific dimensions—such as AI’s knowledge, autonomy, and computational resources—to explore how AI's impact is shaped by these dimensions. These considerations, absent from \cite{antras2006offshoring}, are crucial for understanding AI’s unique effects.\footnote{Another related contribution is \citeauthor{pancs}'s (\citeyear{pancs}, ch. 5) textbook example, which shows how superintelligent machines—matching the abilities of the most capable humans—eliminate labor income inequality. In contrast, we demonstrate that AI’s effects on labor income distribution are nuanced, depending critically on AI's proficiency and autonomy.}

With regard to the existing literature on automation, the first important recent contribution is \cite{zeira1998workers}, who shows how automation can lead to a decline in the labor share as the economy develops. \cite{acemoglu2018race} contend that, by depressing wages, automation also encourages the creation of new tasks in which labor has a comparative advantage. Further developing these ideas, \cite{acemoglu2022tasks} and \cite{acemoglu2024tasks} study how different types of technological advances---such as automation and labor-augmenting technologies---affect wages, inequality, and productivity. Complementary research by \cite{autor2003} and \cite{acemoglu2011chapter} emphasizes how the automation of codifiable tasks—driven by the adoption of computers and robots—explains employment and wage polarization, as these tasks are predominantly performed by middle-skilled workers.\footnote{Other important contributions include \cite{aghion2017artificial}, \cite{moll2022uneven}, \cite{korinek}, \cite{acemoglu2022automation}, \cite{acemoglu2024simple}, and \cite{jones2024AI}. } 

Our work contributes to this literature by focusing on AI, a fundamentally new form of automation. Unlike earlier technologies that primarily automated codifiable tasks, AI can acquire and apply tacit knowledge, allowing it to perform cognitive, non-codifiable work. As discussed above, by automating such work, AI has the potential to ease knowledge bottlenecks and enable a fundamental reorganization of production.

Traditional models of automation, while useful for understanding earlier automation waves, have a hard time capturing this transformation. These models often overlook firms' central role in structuring production—an essential factor for assessing AI’s impact on the knowledge economy. In contrast, our framework explicitly accounts for the organization of knowledge work, showing how AI reshapes production and how shifts in firm composition create new pathways for AI to influence labor outcomes.\footnote{For instance, we show that in the case of a basic autonomous AI, the most knowledgeable human solvers switch toward using AI for routine production work. As a result, a lower-quality pool of solvers is left to assist those who continue as workers in the post-AI world, reducing their productivity. As discussed in Section \ref{sec:displacement}, this outcome is consistent with anecdotal evidence from sectors such as investment banking and legal services.} Moreover, we introduce AI-specific dimensions absent from existing models, leading to new insights. For example, we find that non-autonomous AI disproportionately benefits the least knowledgeable individuals, while autonomous AI primarily benefits the most knowledgeable. As far as we know, these results have no parallel in the existing literature on automation or AI.

Finally, we also contribute to the broader literature on the economics of knowledge and ideas \citep[e.g.,][]{romer1990endogenous,romer1993,jonesgrowth}, the knowledge-based view of the firm \citep[e.g.,][]{nonaka,grant,garicano2000hierarchies,garicanowu2012}, and the economics of data \citep[e.g.][]{tonetti,veldkamp,farboodi}. Specifically, we highlight how modern AI systems blur the conventional boundary between codifiable and non-codifiable knowledge.

Both types of knowledge are, in principle, non-rival—one person’s use does not reduce availability for others---but their practical application depends on rival resources such as time, effort, or computational power. The key distinction between codifiable and non-codifiable knowledge lies in their transferability. Codifiable knowledge, such as information stored in databases or manuals, is highly transferable because it can be expressed as explicit rules and procedures. In contrast, non-codifiable knowledge has limited transferability, spreading primarily through direct experience. Consequently, individuals who possess tacit knowledge often become critical bottlenecks in its application.

Modern AI systems do not fit neatly into this classification. These systems learn from examples, identifying patterns that cannot always be reduced to explicit, formal descriptions. Once trained, however, they can apply these insights at scale using compute. In this sense, AI effectively encodes tacit-like knowledge into model parameters—making it codifiable to machines but not humans. This creates a hybrid form of “embodied” knowledge that is neither purely human nor strictly codified, challenging conventional wisdom on how knowledge is transferred, scaled, and applied.

\subsection*{Roadmap}

The rest of the paper is organized as follows. Section \ref{sec:motivation} documents three key developments in modern AI that shape the assumptions of our baseline model, introduced in Section \ref{sec:model}, which focuses on autonomous AI. Section \ref{sec:characterization} characterizes the equilibrium with autonomous AI, while Section \ref{sec:effects} examines the impact of this technology relative to the pre-AI equilibrium. Section \ref{sec:nonautonomy} explores the implications of non-autonomous AI. Finally, Section \ref{sec:conclusion} concludes by discussing additional implications of our analysis and linking our findings to emerging empirical evidence about AI’s effects on the labor market. We relegate all formal mathematical proofs to the Online Appendix.

\section{Three Key Developments in Modern AI} \label{sec:motivation}

In this section, we document three key developments in modern AI that are essential for understanding the excitement and concerns surrounding its impact on labor and organizational outcomes: (i) AI's ability to scale the use of tacit knowledge, (ii) the advent of foundation models, and (iii) the rise of AI agents capable of performing cognitive work autonomously. These three developments guide the assumptions of our baseline model in Section \ref{sec:model}. 

\vspace{2mm}

\noindent \textit{1. AI Can Use Tacit Knowledge at Scale}.--- AI represents a fundamentally new form of automation. Unlike earlier technologies, AI—like humans—can acquire tacit knowledge, enabling it to perform cognitive, non-codifiable work. As \citet[][p. 7]{autor2024applying} observes: \begin{quote} Pre-AI, computing’s core capability was its faultless and nearly costless execution of... procedural tasks. Its Achilles’ heel was its inability to master... tasks requiring tacit knowledge. Artificial Intelligence’s capabilities are precisely the inverse [...] Rather than relying on hard-coded procedures, AI learns by example, gains mastery without explicit instruction and acquires capabilities that it was not explicitly engineered to possess [...] Like a human expert, AI can weave formal knowledge (rules) with acquired experience to make — or support — one-off, high-stakes decisions. \end{quote}

However, unlike humans—who can only apply their tacit knowledge within the limits of their own time—firms can leverage AI across all available compute in the economy. This reflects the nonrival nature of digital information, which has a negligible marginal cost of reproduction \citep{machineage,tucker}. As  \citet[][p. 177]{waisberg2021ai} explain: \begin{quote} You can ask a question at a seminar or webinar, but you can only get the data as provided by the host or speaker, and a given speaker can only be in one seminar at a time... [In contrast, AI] scales; no longer are you limited by the time of a single domain expert. If their knowledge and behavior are encoded into an AI, you can have as many copies of that AI working at the same time as you want. \end{quote}

\vspace{0.5mm}

\noindent \textit{2. The Advent of Foundation Models}.--- Foundation models represent a transformative shift in AI. These are models characterized by their ability to handle a wide variety of tasks through training on large-scale datasets using self-supervised learning techniques \citep{bommasani2022,CEA}. Models like GPT-4, Gemini, and Claude Sonnet exemplify this approach, serving as flexible, general-purpose platforms that can be adapted to diverse downstream tasks with minimal additional training. This adaptability enables firms to leverage shared AI models, removing the need for costly development of proprietary, domain-specific systems.\footnote{For instance, Andreessen Horowitz, a venture capital firm, surveyed around 100 Fortune 500 companies across industries such as technology, telecom, banking, healthcare, and energy. The survey found that 94\% of these firms were using existing foundation models, while only 6\% were creating their own specialized models \citep{andreessen3}.} As a result, a small number of powerful AI models can function across multiple domains, driving significant economic and industry-wide impact.\footnote{Using the language of computer scientists, foundation models exhibit \textit{emergence}—the ability to develop unexpected capabilities that enable them to perform tasks beyond their explicit training. Combined with scalability, emergence then leads to \textit{homogenization}, defined as the reliance on a few powerful models across multiple domains \citep{bommasani2022}.}

Foundation models have demonstrated their versatility across various industries, from customer service automation to legal research and financial analysis. For instance, Klarna, a fintech company, automated over 700 customer service roles using a version of OpenAI’s models. In the legal field, Harvey leverages OpenAI’s technology to automate research and document analysis. Similarly, Claude Sonnet powers applications at Lonely Planet for itinerary creation and at Bridgewater Associates for tasks like research, coding, and data visualization.

Moreover, evidence suggests that general-purpose models often outperform specialized AI systems. For instance, ChatGPT—a fine-tuned version of GPT-4—has demonstrated superior performance to BloombergGPT---a Large Language Model (LLM) trained specifically on financial data---in various finance-related tasks \citep{bloomberg}. Similarly, \citet{norican2023} demonstrate that GPT-4 surpasses specialist models across medical benchmarks and argue that this advantage extends to other domains, including accounting, law, nursing, and clinical psychology. This shows that even in fields that traditionally relied on highly specialized models, general-purpose foundation models offer better performance, increased flexibility, and lower costs.

\vspace{2mm}

\noindent \textit{3. The Rise of AI Agents.}---The adaptability, efficiency, and scalability of foundation models are driving a new wave of automation: the rise of AI agents capable of performing cognitive work autonomously. Technology firms such as DeepMind and OpenAI are actively developing and starting to deploy these AI agents. As Dario Amodei, co-founder and CEO of Anthropic, explains \citep{amodei}:\begin{quote} I am not talking about AI as merely a tool to analyze data [...] It does not just passively answer questions; instead, it can be given tasks that take hours, days, or weeks to complete, and then goes off and does those tasks autonomously, in the way a smart employee would, asking for clarification as necessary. \end{quote}

The emergence of autonomous AI agents raises the crucial question of whether current AI models are sophisticated enough for such responsibilities. Sarah Tavel, a general partner at the venture capital firm Benchmark, is optimistic \citep{stebbings1}: \begin{quote} There are certainly use cases that can be automated right now, and we see, we see a lot of them... beyond that, there is no question that as the foundation models improve, we are going to see the ability to take on more and more complex tasks. \end{quote} 

This optimism is grounded in the fact that much of today's economy is driven by knowledge work, an area where AI excels compared to industries reliant on manual labor. For instance, \citet{eloundou} estimate that 46\% of jobs could have more than half of their tasks affected by AI foundation models in the near future. Similarly, \citet{goldman} projects that two-thirds of U.S. occupations are at least partially exposed to AI-driven automation, with approximately a quarter of all work potentially automatable when weighted by employment share.

These developments call for new research into AI’s economic implications. As \cite{hadfield} remarked during a panel at the ASSA 2025 meetings in San Francisco:\begin{quote}
All of this is about autonomy […] Think about what that means [...] [AI agents would be] engaging in transactions, entering into contracts, hiring people, designing products, setting prices for products, advertising, doing all of those things. So I think it’s really important for us to be shifting our thinking a little bit. [...] Not just thinking about it as a technology or a tool but rather as a novel economic agent.\end{quote}

\section{Autonomous AI} \label{sec:model}

\subsection{The Model} \label{sec:setting}

\noindent \textit{The Pre-AI Economy}.--- There is a unit mass of humans, each endowed with one unit of time and exogenous knowledge $z \in[0,1]$. The distribution of knowledge in the population is given by a cumulative distribution function $G$ with density $g$. The only assumption we impose on this distribution is that $g$ is continuous and strictly positive for all $z$. The knowledge of each individual is perfectly observable. 

There is a large measure of identical competitive firms. Production occurs inside firms, which are the residual claimants of all output. Labor and knowledge are the sole inputs in production. Firms have no fixed costs and, for simplicity, two layers at most.\footnote{In the Online Appendix, we show that many insights we uncover---such as the conditions under which the least and most knowledgeable individuals benefit from AI---naturally generalize to settings with an arbitrary number of layers.} 

Single-layer firms rely on a single individual to carry out production. This ``independent producer'' dedicates her entire time to pursuing a single production opportunity. These opportunities can be thought of as work requiring cognitive ability, such as handling customer support, drafting contracts, hiring employees, and designing products. Each production opportunity is linked to a problem whose difficulty $x$ is ex-ante unknown and distributed uniformly on $[0,1]$, independently across problems. If the knowledge of the human engaging in production exceeds the problem's difficulty, she solves the problem and produces one unit of output. Otherwise, no output is produced.\footnote{The assumption that $x \sim U[0,1]$ is without loss, as the distribution of knowledge in the population is arbitrary. Under this assumption, an individual's knowledge $z$ is interpreted as the fraction of problems she can solve on her own.}

Two-layer firms hire one ``solver'' and multiple ``workers,'' where all workers have the same knowledge. The restriction that all workers share the same knowledge is without loss because---as we show below---the equilibrium matching arrangement between (human) workers and (human) solvers exhibits strict positive assortative matching. 

As in single-layer firms, each worker in a two-layer firm devotes her time to a single production opportunity. The difference is that if the worker cannot solve the problem independently, she can ask the solver for help. If the solver's knowledge exceeds the problem's difficulty, she communicates the solution to the corresponding worker, who then produces a unit of output. Otherwise, the worker fails to produce. However, communication is costly in that each request for help consumes \( h \in (0,1) \) units of the solver’s time, regardless of whether the solver knows the solution. Hence, a two-layer organization optimally hires exactly $n(z)$ workers of knowledge $z<1$ to fully exploit its solver's time, where $n(z)$ satisfies $h \times n(z) \times (1-z) =1$. 

\begin{figure}[t!]
\centering
\includegraphics[scale=0.68]{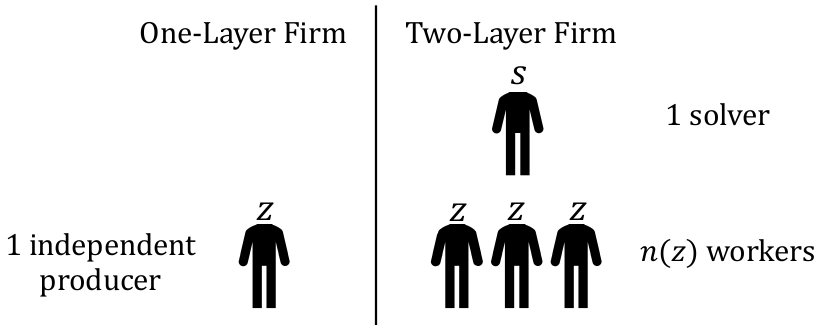}
\captionsetup{justification=centering}
 \caption{The Two Possible Firm Configurations in the Pre-AI World} \label{fig:jungle1}
\end{figure}

Figure \ref{fig:jungle1} depicts the two possible firm configurations of the pre-AI world, where the letter attached to each human corresponds to her knowledge. In this context, knowledge is tacit because solvers cannot prescribe a plan of action to workers in advance, and a worker must interact with her solver to determine whether the solver knows the solution to a problem.

\vspace{2mm}

\noindent \textit{Autonomous AI}.---We model AI as a technology that converts compute into "AI agents," all with the same exogenously fixed level of knowledge, \( \zbar \in [0, 1) \).\footnote{We assume $\zbar<1$ because the equilibrium has a discontinuity at $\zbar=1$. See the Online Appendix for the case $\zbar=1$.} Each AI agent requires one unit of compute and serves as a perfect substitute for a human with equivalent knowledge. These AI agents are “autonomous” because they can function both as “co-workers” (pursuing production opportunities) and as “co-pilots” (providing advice). In Section \ref{sec:nonautonomy}, we explore the case where the AI agents can only be co-pilots, rendering them non-autonomous. 

All firms have access to AI. Post-AI, firms decide their organizational structure and whether to integrate AI agents into their production processes. The total amount of compute in the economy, denoted by \( \mu \), is exogenous. We further assume that production opportunities exceed the combined capacity of humans and AI agents to pursue them, meaning opportunities are abundant relative to available resources. As we discuss below, this implies that there is no unemployment in equilibrium.

\begin{figure}[t!]
\centering
\includegraphics[scale=0.6]{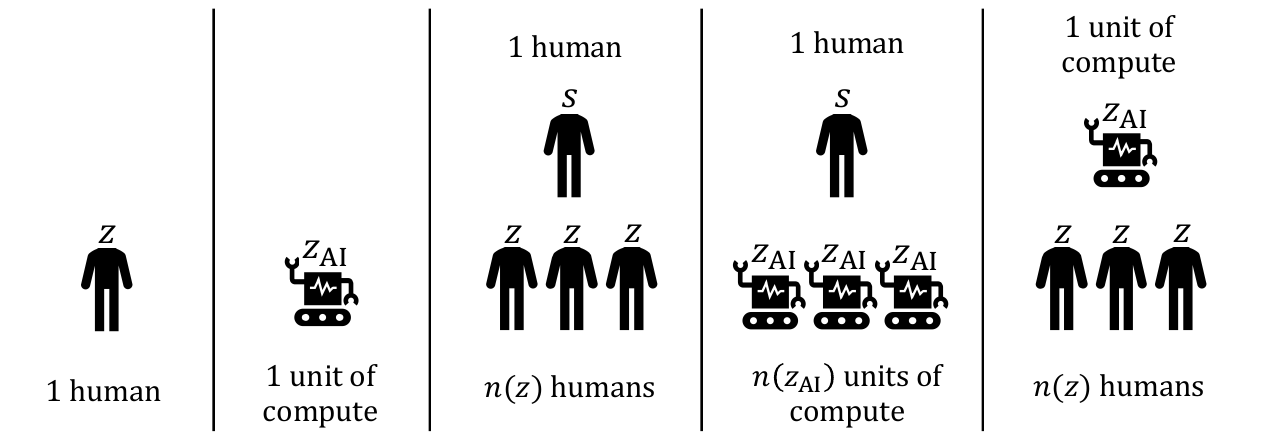}
\caption{The Five Possible Firm Configurations in the Post-AI World} 
\label{fig:jungle2}
\end{figure}

Figure \ref{fig:jungle2} illustrates the five possible post-AI firm configurations. In addition to single- and two-layer firms that hire only humans (the only possible pre-AI firms), there are three additional possibilities: Single-layer automated firms (which use AI agents as independent producers), bottom-automated firms (which use AI agents as workers), and top-automated firms (which use AI agents as solvers). Note that a firm will never use AI agents in both layers of the organization because an AI solver knows the solution to the same set of problems as an AI worker.

Our modeling approach is motivated by the three AI developments described in Section \ref{sec:motivation}. First, unlike humans—who can only apply their knowledge within the limits of their own time—AI is scalable, meaning that once knowledge \( \zbar \) is encoded into the technology, such knowledge can be used simultaneously across all available compute. Second, foundation models—flexible and general-purpose—allow firms to access AI on equal terms through a shared AI model. Third, firms can combine AI with compute to deploy autonomous AI agents capable of performing cognitive work independently.

\vspace{2mm}

\noindent \textit{Wages, Prices, and Profits}.---Let $w(z)$ be the wage of a human with knowledge $z$ and denote by $r$ the rental rate of one unit of compute (i.e., the rental rate of an AI agent). All humans in this economy are risk neutral and maximize their income. In particular, the owners of compute rent it to the highest bidder. We normalize the value of each unit of output to one.

The profit of a single-layer organization is:\begin{equation*} \Pi_1 = \begin{dcases}  z-w(z) & \text{if the firm hires a human with knowledge $z$} \\
\zbar - r & \text{if the firm uses an AI agent} \end{dcases} \end{equation*} In other words, it is equal to the expected output of its independent producer minus the hiring cost.

The profit of a two-layer organization, in turn, depends on whether it uses AI as a solver (i.e., automates the top layer, a ``$t$A'' firm), uses AI as a worker (i.e., automates the bottom layer, a ``$b$A'' firm), or it does not use AI (an ``$n$A'' firm): \begin{equation*} \begin{split} &\Pi_2^{t\mathrm{A}}(z) = n(z)[\zbar-w(z)]-r \qquad \ \ \ \ \ \ \, \text{(where $z \le \zbar$)} \\
 & \Pi_2^{b\mathrm{A}}(s) = n(\zbar)[s-r]-w(s) \qquad \ \ \ \ \ \ \  \text{(where $\zbar \le s$)} \\
&\Pi_2^{n\mathrm{A}}(s,z)=  n(z)[s-w(z)]-w(s) \qquad  \text{(where $z \le s$)} \end{split} \end{equation*}where $z$ and $s$ denote the knowledge of a human worker and a human solver, respectively, and we use the fact that no two-layer firm hires a solver who is less knowledgeable than its workers. In all three cases, the profit of a firm is its expected output minus the cost of the resources it uses. For instance, in the case of a $t$A firm that hires workers with knowledge $z$, its total expected output is $n(z)\zbar$, while the cost of resources is $n(z)w(z)+r$.

 \vspace{2mm}
 
 \noindent \textit{Competitive Equilibrium}.--- A competitive equilibrium comprises the following objects: (1) an allocation of compute across its various potential uses; (2) a partition of the human population into distinct occupations; (3) a description of the matching between different types of agents; and (4) a wage schedule, $w(z)$, and a rental rate of compute, $r$.

Let $\mu_{i}$, $\mu_w$, and $\mu_s$ be the amount of compute used to create AI independent producers, workers, and solvers, respectively. We denote by $I$ the set of humans hired as independent producers, and by $W_p$ and $W_a$ the set of human workers assisted by humans and AI agents, respectively.\footnote{We use the subscript ``$p$'' (for people) instead of ``$h$'' (for humans), to avoid any confusion with the helping cost $h$.} Similarly, let $S_p$  be the set of humans who assist other humans, and $S_a$ the set of humans assisting AI agents in production. We assume that the sets \( I, W_p, W_a, S_p \), and $S_a$ are measurable with respect to $G$. Since a $t$A firm that hires human workers with knowledge $z \in W_a$ rents one AI agent, $\mu_s = \int_{W_a} h(1-z) dG(z)$. Similarly, since a $b$A firm that hires a human solver with knowledge $s \in S_a$ rents $n(\zbar)$ AI agents, $\mu_w=n(\zbar) \int_{S_a} dG(z)$. 

Finally, let \( m: W_p \to S_p \) represent the function describing the pointwise matching arrangement generated by the hiring decisions of $n$A firms. Specifically, $m(z)$ denotes the knowledge of the human solver assisting human workers with knowledge $z$.\footnote{As discussed in \cite{fuchs2015optimal}, $m(z)$ is an approximation of a firm that hires a small interval of human workers $(z-\varepsilon_1,z+\varepsilon_1)$ and a small interval of human solvers $(m(z)-\varepsilon_2,m(z)+\varepsilon_2) $ with the requirement that the mass of solvers in the firm $\int_{m(z)-\varepsilon_2}^{m(z)+\varepsilon_2}dG(u)$ is equal to $h$ times the mass of problems left unsolved by the workers in the firm $ \int_{z-\varepsilon_1}^{z+\varepsilon_1}(1-u)dG(u)$. The latter requirement captures that $n$A firms optimally hire $n(z)$ workers of knowledge $z$.} As mentioned earlier, the equilibrium matching arrangement between human workers and human solvers exhibits strict positive assortative matching, so $m$ is strictly increasing. Additionally, it must satisfy the following resource constraint: \begin{equation} \label{eq:budget}\textstyle \int_Y h(1-u) dG(u) = \int_{m(Y)} dG(u), \ \text{for any measurable set $Y \subseteq W_p$} \end{equation}where \( m(Y) \) denotes the set of solvers assigned to the set of workers \( Y \subseteq W_p \). This condition ensures that the total time required to consult on problems left unsolved by the workers in \( Y \) equals the total time available from the solvers matched to these workers, \( m(Y) \). In essence, it reflects that $n$A firms optimally hire \( n(z) \) workers with knowledge \( z \) to fully use the time of their solvers.

\begin{mydef} An equilibrium consists of nonnegative amounts $(\mu_{i},\mu_w,\mu_s)$, measurable sets $(I,W_p,W_a,S_p,S_a)$, a function $m: W_p \to S_p$ satisfying (\ref{eq:budget}), a wage schedule $w: [0,1] \to \mathbb{R}_{\ge 0}$ and a rental rate of compute $r \in \mathbb{R}_{\ge 0}$, such that: \begin{enumerate}[leftmargin=6mm,noitemsep]
\item Firms optimally choose their structure (while earning zero profits).
\item Markets clear: (i) $\mu_{i}+\mu_w+\mu_s=\mu$, and (ii) the union of the sets $(I,W_p,W_a,S_p,S_a)$ is $[0,1]$ and the intersection of any two of these sets has measure zero. 
\end{enumerate} \end{mydef}

Note that the human workers in $W_p$ are endogenously matched with the human solvers in $S_p$ according to the pointwise matching function $m$. In contrast, all the humans in $W_a$ are matched with an AI solver (which has knowledge $\zbar$), while all the humans in $S_a$ are matched with AI workers (whose knowledge is $\zbar$).

Total output in this economy is $\int_{I} z dG(z) + \zbar \mu_i+ \int_{W_p} m(z) dG(z)+ \int_{S_a}  n(\zbar) z dG(z)+\int_{W_a} \zbar dG(z)$, where each term represents the total output of each of the five firm configurations depicted in Figure \ref{fig:jungle2} (in the same order). Total labor income, in turn, is equal to $\int_{0}^1 w(z) dG(z)$, whereas total capital income is $\mu r$. Since firms earn zero profits in equilibrium, total output equals the sum of total labor income and total capital income.

 \vspace{2mm}
 
\noindent \textit{Compute is ``Abundant'' Relative to Human Time}.--- We focus on the case in which compute is sufficiently abundant so that the binding constraint in human-AI interactions is human time, not compute. In other words, there are more AI agents than can be matched with humans, implying that some AI agents must engage in independent production.\footnote{A sufficient condition for this to hold is $\int_{0}^{\zbar}n(z)^{-1}dG(z) + n(\zbar)\big( 1- G(\zbar) \big)  < \mu$. For any distribution $G$ and helping cost $h \in (0,1)$, there exists a finite $\mu$ that satisfies this condition for all $\zbar \in [0,1)$. }

We consider this the most relevant case due to the exponential growth in computational capacity over the past two centuries \citep{nordhaus} and the fact that current AI models process information and generate outputs 10 to 100 times faster than humans \citep{amodei}. However, to explore the effects of this assumption, we also examine the case of limited compute in the Online Appendix.

\vspace{2mm} 
 
\noindent \textit{Some Notation}.--- For future reference, we define $W \equiv W_a \cup W_p$ and $S \equiv S_a \cup S_p$ as the overall set of human workers and solvers of the economy, respectively. We also denote by $e: S_p \to W_p$ the inverse of the matching function $m$. That is, $e$ is the ``employee matching function'' denoting the knowledge of the human worker matched with a human solver with knowledge $s \in S_p$. This function always exists given that, as argued above, the matching function $m$ is strictly increasing in equilibrium.

Finally, for any arbitrary set $B$, we denote by $\mathrm{int} B$ and by $\mathrm{cl} B$ the interior and closure of $B$, respectively. We also use $B \preceq B'$ to indicate that the set $B \subseteq [0,1]$ ``lies below'' the set $B' \subseteq [0,1]$. Formally, $B \preceq B'$ if $\sup B \le \inf B'$. For example, $W_a \preceq W_p$ means that the most knowledgeable human worker assisted by AI is weakly less knowledgeable than the least knowledgeable human worker assisted by another human.

\subsection{Discussion of the Model} \label{sec:real}

Before proceeding with the analysis, we map our model to real-world applications. We also clarify how to interpret compute in our framework, justify our focus on a single AI technology, and explain why we do not consider the possibility of unemployment. The Online Appendix discusses some limitations of our model and suggests avenues for future research.

\vspace{2mm}

\noindent \textit{Interpretation.---} In our benchmark model, AI agents are autonomous in that they can operate both as co-workers---capable of pursuing production opportunities---and as co-pilots---providing assistance.\footnote{The model is silent about where production opportunities come from. The role of organizations is to decide who to assign to routine work and who to assign to specialized problem solving. Hence, workers and independent producers---human or AI---should not be thought of as being responsible for generating production opportunities; they are only responsible for pursuing them.} A real-world example of an AI co-worker is Klarna, which automated over 700 customer service agents using OpenAI’s models (see Section \ref{sec:motivation}). Another is Agentforce by Salesforce, an AI agent that not only identifies business leads using company data but also contacts prospects and schedules meetings on behalf of salespeople \citep{rosenbush}. Similarly, All Day TA, a startup founded by  University of Toronto Professors Kevin Bryan and Joshua Gans, leverages foundation models to fully automate the work of university teaching assistants.

In contrast, \cite{brynjolfsson2023generative} provide an example of an AI co-pilot in the customer support industry: their study highlights the productivity effects of equipping human agents with a real-time, AI-based conversational assistant to manage customer inquiries. Another example is GitHub Copilot, which aids developers by suggesting code snippets, autocompleting functions, and recommending improvements to code quality. Similarly, BrainTrust AIR assists HR professionals by reviewing candidates, grading applications, and generating detailed scorecards to streamline recruitment decisions.
 
Regarding AI autonomy, for simplicity, we focus on the two extreme cases. First, we consider a scenario in which AI agents are fully autonomous and capable of performing all tasks that humans possessing knowledge $\zbar$ can accomplish. Then, in Section \ref{sec:nonautonomy}, we examine the opposite case, where AI agents are entirely non-autonomous and can only provide advice. Most AI applications, however, operate between these extremes, with autonomy levels varying by sector. For example, tools like Harvey automate many routine legal tasks, such as document review and contract analysis. Yet, they remain unable to engage directly with clients or argue cases in court.

Because our framework models the broader knowledge economy, we interpret our results as approximations of AI’s impact based on its average level of autonomy across the different relevant sectors. Accordingly, the practical implications of our analysis will likely fall between the predictions of these two benchmark cases. To determine which extreme better reflects current and future realities, the critical question is: How many tasks traditionally performed by humans can AI agents like Harvey handle autonomously, and how will this capability evolve over time?

\vspace{2mm}

\noindent \textit{On Compute.---} Our main goal is to analyze how AI affects human labor outcomes. For this reason, we take AI technology as given. Thus, in our framework, ``compute'' refers to the resources required during the inference phase---the stage in which an AI system generates outputs based on new input data after it has already been trained. 

To facilitate comparison between AI output and human labor, we express compute in terms of human time. In practice, however, compute is often measured in FLOPs or, for LLMs, in tokens. In an earlier version of this paper \citep{idetalamasAI}, we explain how compute in our model translates into these real-world units.

\vspace{2mm}

\noindent \textit{Why Only One AI?---} A key property of AI is that it relies on compute, which is general-purpose hardware that can be rented on demand. This implies that competition among AI providers drives compute toward the highest-performing AI systems—a phenomenon that computer scientists and venture capitalists call the commoditization of foundation models \citep{navani,stebbings3}.\footnote{Compute can be rented on demand through cloud computing providers and easily redeployed across various AI foundation models. According to a recent survey by Andreessen Horowitz, ``Most enterprises are designing their applications so that switching between models requires little more than an API change. Some companies are even pre-testing prompts so the change happens literally at the flick of a switch'' \citep[see][]{andreessen3}.} As Aravind Srinivas, founder of PerplexityAI, explains \citep{stebbings3}:\begin{quote} [It] is a losing game... every time you end up finishing a large training run, you burn a lot of money, you have a great model, and then you watch it be destroyed by the next update... So where are you recovering all that money back? (...) Nobody wants to use the API [i.e., the model] if somebody else is offering a better model at a cheaper price. \end{quote}

This competition for compute underpins our assumption of focusing on a single AI technology. Indeed, in the Online Appendix, we show that if multiple AI technologies are available and all of them require similar compute to create AI agents, only the most advanced AI will be used in equilibrium. However, if less advanced AI technologies require substantially less compute, they may coexist alongside more advanced models. This result is consistent with the development of smaller foundation models, such as GPT-4o Mini and Claude 3 Haiku.

Furthermore, this extension highlights a fundamental distinction between human and AI heterogeneity. While humans with varying levels of knowledge can coexist in equilibrium despite requiring the same resources—time—to perform knowledge work, AI heterogeneity emerges only when models differ significantly in both knowledge and resource requirements. Hence, introducing a population of AI agents is fundamentally different from expanding the human labor force.

\vspace{2mm}
 
 \noindent \textit{No Unemployment}.--- In defining our competitive equilibrium, we excluded the possibility of unemployment. This is without loss of generality, as we assume that total available resources—human time and compute—are scarce relative to production opportunities (even though compute is abundant relative to time). This scarcity ensures that, in equilibrium, it is always worthwhile for every human to be employed in some capacity, even if AI is more knowledgeable than a fraction of the human population. 

In the Online Appendix, we explore an extension where compute is so abundant that it outstrips production opportunities. In that case, we show that AI leads to technological unemployment: All humans who are less knowledgeable than AI become unemployed. Organizations, however, still display a hierarchical structure in that the most knowledgeable humans specialize in tackling the problems that AI cannot solve.

\subsection{Benchmark: The Pre-AI Equilibrium} \label{sec:bench}

We begin by presenting a partial characterization of the equilibrium without AI. This is also the equilibrium when the amount of compute $\mu$ is zero, and was originally described by \cite{fuchs2015optimal}.\footnote{An economy with $\mu=0$ is different from one with $\mu>0$ and $\zbar=0$. This is because, even if AI cannot solve any problems, it can still pursue production opportunities, thereby expanding the economy's production possibility frontier.} Note that in this case, $W_a=S_a=\emptyset$, so $W=W_p$ and $S=S_p$.

\begin{prop} \label{prop:noAI} In the absence of AI, there is a unique equilibrium. The equilibrium is efficient (i.e., it maximizes total output) and has the following features: \begin{itemize}[leftmargin=*,noitemsep]
\item Occupational stratification: $ W \preceq I \preceq S$.
\item Positive assortative matching: The function $m: W \to S$ is strictly increasing.
\item $W \neq \emptyset$ and $S \neq \emptyset$. However, $I \neq \emptyset$ if and only if $h > h_0 \in (0,1)$.\end{itemize} Moreover, the wage function $w$ is continuous, strictly increasing, and convex (strictly so when $z \in W \cup S$), and is given by:
 \begin{itemize}[leftmargin=*,noitemsep] 
\item $w(z)= m(z)-w(m(z))/n(z)$ for all $z \in W$.
\item $w(z)=z$ for all $z\in I$.
\item $w(z) = C+ \int_{\inf S}^z n(e(u))du> z$ for all $z \in S$, where $C > \inf S$ when $h < h_0$ (and $C =  \inf S$ otherwise).
\end{itemize}In particular, $w(z)>z$ for all $z \notin \mathrm{cl} I$ (so $w(z)>z$ for all $z\in [0,1]$ when $h < h_0$). \end{prop}

 \begin{figure}[t!]
  \centering
  \begin{tikzpicture}[xscale=6.8, yscale=4]
 \draw[->] (-.02, 0)--(1.03,  0);
\draw[->] (0, -.02)--(0,1.7);
\draw[dotted,thick] (0,0)--(1,1);
\node[below] at (.9,.9) {$45^{\circ}$}; 
\node[right] at (1.03,0) {\Large $z$}; 
\draw[ultra thick, domain=0:{3-sqrt(5)}, samples=200] plot (\x, {3-sqrt(5)- (1/2*(3-sqrt(5))*(1+1/4*(3-sqrt(5))))/(1+1/2-1/2*(3-sqrt(5))) *(1-\x)  
+1/2*\x^2/2});
\draw[ultra thick, domain={3-sqrt(5)}:1, samples=200] plot (\x, {(1/2+2*(3-sqrt(5)))/(1+1/2*(1-(3-sqrt(5))))-sqrt(1+2*(3-sqrt(5))*2-2*\x*2)});
\draw[dashed] ({3-sqrt(5)},0)--({3-sqrt(5)},{3-sqrt(5)- (1/2*(3-sqrt(5))*(1+1/4*(3-sqrt(5))))/(1+1/2-1/2*(3-sqrt(5))) *(1-(3-sqrt(5)))  +1/2*(3-sqrt(5))^2/2}) ;
\draw (1,-0.02)--(1,0.02); 
\draw[dotted] (1,0)--(1, 1.55);
\node[below] at (0,0) {$0$}; 
\node[below] at (1,0) {$1$}; 
\node[left] at (0,1.5778) {$1.58$}; 
\node[left] at (0,0.35702) {$0.36$}; 
\draw[dotted] (0,1.5778)--(1,1.5778);
\node[above] at (0.85,1.2) {$w(z)$}; 
\draw [decorate, decoration={brace, amplitude=10pt}] ({3-sqrt(5)},0)--(0,0);
 \node[below] at ({(3-sqrt(5))/2},-.1) {$W$}; 
\draw [decorate, decoration={brace, amplitude=10pt}] (1,0) -- ({3-sqrt(5)},0);
 \node[below] at ({3-sqrt(5)+(1-(3-sqrt(5)))/2+0.01},-.1) {$S$}; 
 \draw[thick, densely dashed, <->, out=60, in=120] (.1,0) to (.81,0);
   \draw[thick, densely dashed, <->, out=60, in=120] (.45,0) to (.93,0); 
\fill (.1,0) circle[radius=.15pt];
\fill (.45,0) circle[radius=.15pt];
\node[above] at ({1/2},-0.5) { (a) Pre-AI Equilibrium when $h < h_0$};  

\draw[->] (-0.02+1.25, 0)--(1.05+1.25 ,  0);
\draw[->] (0+1.25, -.02)--(0+1.25,1.7);
\draw[dotted,thick] (0+1.25,0)--(1+1.25,1);
\node[below] at (0.9+1.25,.9) {$45^{\circ}$}; 
\node[right] at (1.05+1.25,0) {\Large $z$}; 
\draw (1+1.25,-0.02)--(1+1.25,0.02); 
\draw ({1.25+0.56},-0.02)--({1.25+0.56},0.02); 
\draw ({1.25+0.67},-0.02)--({1.25+0.67},0.02); 
\draw[dotted] (1+1.25,0)--(1+1.25, { 1.673-sqrt(2.657-2.4615*(2.2-1.2))});
\node[below] at (0+1.25,0) {$0$}; 
\node[below] at (1+1.25,0) {$1$}; 
\node[left] at (0+1.25,1.23) {$1.23$}; 
\node[left] at (0+1.25,0.1262459024) {$0.13$}; 
\draw[dotted] (0+1.25,1.23)--(1+1.25,1.23);
\node[above] at (0.8+1.25,0.94) {$w(z)$}; 
\draw[dashed] ({1.25+0.56},0)--({1.25+0.56},0.56);
\draw[dashed] ({1.25+0.67},0)--({1.25+0.67},.67);
\draw [decorate, decoration={brace, amplitude=10pt}] ({1.25+0.67},0)--({1.25+0.56},0);
 \node[below] at ({1.25+0.57+(.67-.56)/2},-.10) {$I$}; 
 \draw [decorate, decoration={brace, amplitude=10pt}] ({1.25+0.56},0)--(1.3,0);
 \node[below] at ({1.25+0.56/2},-.1) {$W$}; 
 \draw [decorate, decoration={brace, amplitude=10pt}] (1+1.25,0) -- ({1.25+0.67},0);
 \node[below] at ({1.25+0.67+(2.22-(1.25+0.67))/2},-.1) {$S$}; 
  \draw[thick, densely dashed, <->, out=60, in=120] ({1.25+.1},0) to ({1.25+.75},0);
   \draw[thick, densely dashed, <->, out=60, in=120] ({1.25+.45},0) to ({1.25+.96},0); 
    \fill ({1.25+.1},0) circle[radius=.15pt];
      \fill ({1.25+.45},0) circle[radius=.15pt];
\draw[ultra thick, domain=1.25:{1.25+0.56}, samples=200] plot (\x, {(\x-1.25) + .406*(.5575-(\x-1.25))^2});
\draw[ultra thick] ({1.25+0.56},.56)--({1.25+0.67}, 0.67);
\draw[ultra thick, domain={1.25+0.67}:2.25, samples=200] plot (\x, { 1.673-sqrt(2.657-2.4615*(\x-1.25))}); 
\node[above] at ({1/2+1.25},-0.5) {  (b) Pre-AI Equilibrium when $h > h_0$};  
  \end{tikzpicture}
  \captionsetup{justification=centering}
\caption{Illustration of the Pre-AI Equilibrium. \\  \justifying 
 \vspace{0.5mm}
\footnotesize \noindent \textit{Notes.} Distribution of knowledge: $G(z)=z$. Parameter values: For panel (a), $h=1/2$ ($<h_0= 3/4$), while for panel (b), $h=0.8125$ ($>h_0= 3/4$). The thick line depicts the equilibrium wage function. The dashed arrows illustrate the matching between workers and solvers.}
\label{fig:pre}\justifying
\end{figure}

The equilibrium without AI---which we illustrate in Figure \ref{fig:pre}---has several salient features.  First, it is efficient, as the First Welfare Theorem holds in this economy (with perfect competition, complete information, and only pecuniary externalities). Moreover, the wage of an individual with knowledge $z$ corresponds to her marginal product, defined as the increase in output from introducing one additional human with this knowledge into the economy.

Second, the equilibrium exhibits occupational stratification: Workers are less knowledgeable than independent producers, who are, in turn, less knowledgeable than solvers. Intuitively, more knowledgeable individuals have a comparative advantage in specialized problem solving, as this allows them to leverage their knowledge by applying it to more than one problem. Hence, $ (W \cup I) \preceq S$. Similarly, less knowledgeable individuals have a comparative advantage in assisted rather than independent production work, as they are less likely to succeed on their own. Consequently, $W \preceq  I$.

Third, there is strict positive assortative matching: More knowledgeable workers in $W$ match with more knowledgeable solvers in $S$. The reason is that worker and solver knowledge are complements: For a given team of workers, a more knowledgeable solver increases expected output, while for any given solver, more knowledgeable workers increase team size. 

Fourth, the set of workers and the set of solvers are always nonempty, but the set of independent producers is empty when the communication cost $h$ is below a threshold $h_0 \in (0,1)$. Intuitively, lower values of \(h\) make two-layer organizations more attractive than single-layer ones, leading all humans to be employed by two-layer organizations. 

Fifth, workers and solvers earn strictly more than their expected output as independent producers (except in the case of the most knowledgeable worker and the least knowledgeable solver when $h \ge h_0$). The reason for this is that the marginal value of knowledge is strictly lower than $1$ for workers (as their knowledge is used to free up solver time),\footnote{A marginal increase in the knowledge of a worker with knowledge $z$ frees $h<1$ units of her solver's time. This allows her firm to hire $1/(1-z)$ extra workers, with an expected net output gain of $(m(z)-w(z))/(1-z)<1$ (as the wage $w(z)$ of that worker is strictly greater than $z$ in the interior of $W$).} exactly equal to $1$ for independent producers (as their expected output equals their knowledge), and strictly greater than $1$ for solvers (as they can use their knowledge in more than one problem). Hence, if solvers earned their expected output as independent producers, then all two-layer firms would try to hire the most knowledgeable individuals as solvers. Similarly, if workers earned their expected output as independent producers, then all two-layer firms would try to hire the least knowledgeable individuals as workers.

Finally, the equilibrium wage function $w$ is continuous, strictly increasing, and convex (strictly so when $z \in W \cup S$). The wages for the different occupations are obtained as follows. For independent producers, $w(z)=z$ is an immediate implication of the zero-profit condition of single-layer firms. For workers and solvers, consider the problem of a two-layer organization that recruited $n(z)$ workers with knowledge $z \in W$ and is deciding which solver $s \in S$ to hire: \[ \max_{s \in S} \Big\{ n(z)[s-w(z)]-w(s) \Big\}  \]The corresponding first-order condition evaluated at $s=m(z)$ implies that $w^{\prime}(m(z))=n(z)$, or, equivalently, $w^{\prime}(z)=n(e(z))$ for any $z \in S$. Thus $w(z) =C+ \int_{\inf S}^z n(e(u))du$ for any $z \in S$, where the constant $C$ is chosen so that the wage function is continuous, as the latter condition is necessary for market clearing. The wages of workers are then determined by the zero-profit condition of two-layer organizations: $w(z) = m(z) - w(m(z))/n(z)$ for any $z \in W$.

For simplicity, we restrict attention to $h< h_0$. This implies that there are no independent producers in the pre-AI equilibrium. In a previous version of this paper \citep{idetalamasAI_old}, we show that virtually all of our results---except one, discussed in Section \ref{sec:wages}---extend to the case where $h \ge h_0$.

\section{The Equilibrium with Autonomous AI}\label{sec:characterization}

Our objective is to understand the effects of autonomous AI by comparing the pre- and post-AI equilibrium. In this section, we provide a partial characterization of the post-AI equilibrium, focusing on the essential information needed for the comparison (which is carried out in Section \ref{sec:effects}).

For future reference, we index the post-AI equilibrium using the superscript ``$\post$'' (note that the pre-AI equilibrium has no superscript). Furthermore, recall that $W^* \equiv W_a^* \cup W_p^*$ is the overall set of human workers and that $S^* \equiv S_a^* \cup S_p^*$ is the overall set of human solvers.

\begin{prop} \label{prop:eqAI} In the presence of an autonomous AI, there is a unique equilibrium. The equilibrium is efficient and has the following features:
\begin{itemize}[leftmargin=*,noitemsep]
\item Occupational stratification: $ W^* \preceq I^* \preceq S^*$.
\item No worker is more knowledgeable than AI, and no solver is less knowledgeable than AI: $ W^* \preceq \{\zbar\} \preceq S^* $.
\item Positive assortative matching: $m^*: W_p^* \to S_p^*$ is strictly increasing and $W_a^* \preceq W_p^*$ and $S_p^* \preceq S_a^*$.
\end{itemize} 
Furthermore, AI is always used for independent production, and whether it is also used as a worker or as a solver depends on its knowledge level \textbf{relative to the pre-AI equilibrium}:  
\begin{itemize}[leftmargin=*,noitemsep]
\item If $\zbar \in W$, then AI is necessarily used as a worker (and possibly also as a solver).
\item If $\zbar \in S$, then AI is necessarily used as a solver (and possibly also as a worker).
 \end{itemize}
Finally, the rental rate of compute $r^*$ is equal to $\zbar$, and the wage function $w^*$ is continuous, strictly increasing, and convex (strictly so when $z \in W_p^* \cup S_p^*$), and is given by: \begin{itemize}[leftmargin=*,noitemsep] 
\item $w^{\post}(z)=\zbar(1-1/n(z))>z$ for all $z \in W_a^{\post}$.
\item $w^{\post}(z)= m^*(z)-w^{\post}(m^*(z))/n(z)$ for all $z \in W_p^{\post}$.
\item $w^{\post}(z)=z$ for all $z\in I^{\post}$.
\item $w^{\post}(z) = C^*+ \int_{ \inf S_p^{\post}}^z n(e^*(u))du$ for all $z \in S_p^{\post}$, where $C^*=\inf S_p^{\post}$.
 \item $w^{\post}(z)=n(\zbar)(z-\zbar)>z$ for all $z \in S_a^{\post}$.
\end{itemize}In particular, $w^*(\sup W^*)=\sup W_p^*$, $w^*(\zbar)=\zbar$, and $w^*(\inf S^*)=\inf S_p^*$.
\end{prop}

Figure \ref{fig:postAI} illustrates the post-AI equilibrium in two scenarios. In panel (a), AI agents serve only as workers and independent producers but not as solvers. In panel (b), AI agents perform all three roles. In both cases, AI agents are the most knowledgeable workers, so they are assisted by the most knowledgeable human solvers. In panel (b), AI agents are also the least knowledgeable solvers, so they assist the production work of the least knowledgeable human workers. Regardless of the scenario, humans with knowledge $\zbar$ earn their output as independent producers because they are perfect substitutes for AI agents (each priced at \(r^* = \zbar\)).

\begin{figure}[!b]
    \centering
  \begin{tikzpicture}[xscale=5.8140, yscale=3.42]
\draw[->] (-.02, 0)--(1.03,  0);
\draw[->] (0, -.02)--(0,2.05);
\draw[dotted,thick] (0,0)--(1,1);
\node[right] at (1.03,0) {\Large $z$}; 
\draw (1,-0.02)--(1,0.02); 
\node[below] at (0,0) {$0$}; 
\node[below] at (1,0) {$1$}; 
\node[left] at (0,2) {$2$}; 
\draw[dotted] (0,2)--(1,2);
\draw[dashed] (0.425, 0)--(0.425, 0.425); 
\draw[dotted] (1, 0)--(1,2);
\draw[dotted] (1, 2)--(1,2+0.05);
 \node[left] at (0,0.425) {$\zbar$};
  \node[left] at (0, 0.2566) {$0.267$};
\draw[dashed] (0, 0.425)--(0.425, 0.425); 
\node[below] at (0.425,-.04) {$\zbar$};
\node[above,black] at (.80,1.5777) {$w^*(z)$}; 
\draw[ultra thick, domain=0:0.426, samples=100,black] plot (\x+0, {0.26656 + 0.26656*\x + 0.25000*\x^2});
\draw[ultra thick,black] (0.425+0,0.425)--(0.53311+0,0.53311);
\draw[ultra thick, domain=(0.53311-0.001):0.7009, samples=100,black] plot (\x+0, {1.5331 - sqrt(3.1323 - 3.9998*\x)});
\draw[ultra thick, domain=(0.70046-0.001):1, samples=100,black] plot (\x+0, {-1.4783 + 3.4783*\x});

 \draw [decorate, decoration={brace, amplitude=10pt}] (0+0,2.1-0.03)--(0.425+0,2.1-0.03);
 \node[above] at (0.425*.5+0+.005,2.18-0.03) {$W_p^{\post}$}; 
 \draw [decorate, decoration={brace, amplitude=10pt}] (0.425+0,2.1-0.03)--(0.53311+0,2.1-0.03);
 \node[above] at ({0.425+(0.53311-0.425)/2+0+.015},2.225-0.03) {$I^{\post}$}; 
 \draw [decorate, decoration={brace, amplitude=10pt}] (0.53311+0,2.1-0.03)--(0.70046+0,2.1-0.03);
 \node[above] at ({0.53311+(0.70046-0.53311)/2+0+.0075},2.18-0.03) {$S_p^{\post}$}; 
 \draw [decorate, decoration={brace, amplitude=10pt}] (0.70046+0,2.1-0.03)--(1+0,2.1-0.03);
 \node[above] at ({0.70046+(1-0.70046)/2+0+.0075},2.19-0.03) {$S_a^{\post}$}; 
  
 \draw[dotted] (.425, 0.425)--(.425, 2.03+0.05);
\draw[dotted] (0.53311, 0)--(0.53311, 2.03+0.05);
\draw[dotted] (0.70046, 0)--(0.70046, 2.03+0.05);

\node[below] at (.7,.7) {$45^{\circ}$}; 
   \draw[thick, densely dashed, <->, out=60, in=120] (.1,0) to (0.5806140351,0); 
    \draw[thick, densely dashed, <->, out=60, in=120] (.25,0) to (0.6424890351,0);   
    \draw [decorate, decoration={brace, amplitude=15pt}] (1,0)--(0.70046,0);   
       \draw[thick, densely dashed, <->, out=-90, in=-90] (0.425,-0.17) to (0.85,-0.17);    
\node[above] at ({1/2},-0.6) { (a) AI is used as a worker but not as a solver};   

       
  \draw[->] (1.3-.02, 0)--(1.3+1.03,  0);
\draw[->] (1.3+0, -.02)--(1.3+0,2.05);
\draw[dotted,thick] (1.3+0,0)--(1.3+1,1);
\node[right] at (1.3+1.03,0) {\Large $z$}; 
\draw (1.3+1,-0.02)--(1.3+1,0.02); 
\node[left] at (0+1.3,2) {$2$}; 
\node[left] at (0+1.3,0.425) {$0.425$}; 
\draw[dotted] (0+1.3,2)--(1+1.3,2);
\node[below] at (1.3+0,0) {$0$}; 
\node[below] at (1.3+1,0) {$1$}; 
\node[below] at (1.3+0.85,-.04) {$\zbar$};
\node[left] at (1.3,0.85) {$\zbar$};
\draw[dashed] (1.3,0.85)--(1.3+.85,.85);       
\draw[dotted] (1.3+1,0)--(1.3+1,2); 
\draw[dotted] (1.3+1,2)--(1.3+1,2.03+0.05); 
\draw[dashed] (1.3+.85,0)--(1.3+.85,0.13);       
\draw[dashed] (1.3+.85,0.22)--(1.3+.85,0.85);      
\draw[dashed] (1.3+.85,0)--(1.3+.85,0.02);              
 \draw [decorate, decoration={brace, amplitude=10pt}] (0+1.3,2.03+0.05)--(.345025+1.3,2.03+0.05);
 \node[above] at (.345025/2+1.3,2.11+0.05) {$W_a^{\post}$}; 
 \draw [decorate, decoration={brace, amplitude=10pt}] (.345025+1.3,2.03+0.05)--(.85+1.3,2.03+0.05);
 \node[above] at ({.345025+(.85-.345025)/2+1.3},2.11+0.05) {$W_p^{\post}$}; 
  \draw [decorate, decoration={brace, amplitude=10pt}] (.85+1.3,2.03+0.05)--(.948+1.3,2.03+0.05);
 \node[above] at ({.86+(.948-.85)/2+1.3-0.02},2.11+0.05) {$S_p^*$}; 
   \draw [decorate, decoration={brace, amplitude=10pt}] (.948+1.3,2.03+0.05)--(1+1.3,2.03+0.05);
 \node[above] at ({.948+(1-.948)/2+1.3+0.01},2.11+0.05) {$S_a^*$}; 
\draw[ultra thick, domain=0:.345025, samples=100] plot (\x+1.3, {.425+.425*\x});
\draw[ultra thick, domain=.345025:.85, samples=100] plot (\x+1.3, {.45476+.252488*\x+.25*\x^2});
\draw[ultra thick, domain=.85:.953, samples=100] plot (\x+1.3, {1.50498-sqrt(3.82899-4*\x)});
\draw[ultra thick, domain=.951623:1, samples=100] plot (\x+1.3, {-11.3333+13.3333*\x});
\node[above,black] at (1.3+.18,0.52) {$ w^*(z)$};  
\draw[dotted] (0.345025+1.3, 0)--(0.345025+1.3, 2.03+0.05);
\draw[dotted] (.85+1.3, 0.85)--(.85+1.3, 2.03+0.05);
\draw[dotted] (.948+1.3, 0)--(.948+1.3, 2.03+0.05);
\node[below] at (.7+1.3,.7) {$45^{\circ}$}; 
       \draw[thick, densely dashed, <->, out=-90, in=-90] (0.85+1.3,-0.17) to (0.345025/2+1.3,-0.17);    
 \draw [decorate, decoration={brace, amplitude=15pt}] (.345025+1.3,0)--(0+1.3,0);
   \draw [decorate, decoration={brace, amplitude=12pt}] (.948+1.3,0)--(1+1.3,0);
\draw[thick, densely dashed, <->, out=60, in=120] (0.85+1.3,0.15) to (0.974+1.3,0.15);       
\draw[thick, densely dashed, <->, out=60, in=120] (.45+1.3,0) to (0.8816231435+1.3,0);       
\draw[thick, densely dashed, <->, out=60, in=120] (.65+1.3,0) to (0.9266231435+1.3,0);    
\node[above] at ({1/2+1.3},-0.6) { (b) AI is used as a worker and a solver};      

  \end{tikzpicture}
\captionsetup{justification=centering}
 \caption{Illustration of the Post-AI Equilibrium for Two Different Values of $\zbar$ \\ \justifying 
  \vspace{0.5mm}
 \footnotesize \noindent  \textit{Notes}. Distribution of knowledge: $G(z)=z$. Parameter values: Both panels have $h=1/2$. For panel (a), $\zbar=0.425$, while for panel (b), $\zbar=0.85$. The thick line depicts the equilibrium wage function. The dashed arrows illustrate the matching between workers and solvers, both humans and AI. Human workers in $W_p^*$ are endogenously matched with the human solvers in $S_p^*$ according to $m^*$. All the humans in $W_a^*$ are matched with an AI solver, while all humans in $S_a^*$ are matched with AI workers.} \label{fig:postAI}
\end{figure}

Proposition \ref{prop:eqAI} has three parts. The first part states the basic properties of the equilibrium. The second part describes how firms use AI agents as a function of their knowledge. The third part characterizes the equilibrium prices. 

Let us consider part one. As in the pre-AI world, the post-AI equilibrium is unique and efficient, exhibits occupational stratification, and features positive assortative matching. The remaining properties follow from occupational stratification, positive assortative matching, and AI’s scalability. Indeed, because compute is abundant relative to time, some AI agents must engage in independent production. Hence, occupational stratification implies no worker is more knowledgeable than AI, and no solver is less knowledgeable than AI (i.e., $ W^* \preceq \{\zbar\} \preceq S^*$). Moreover, by positive assortative matching, if AI is used as a worker, then it is supervised by the most knowledgeable solvers (i.e., $S_p^* \preceq S_a^*$), and if AI is used as a solver, then it assists the least knowledgeable workers (i.e., $W_a^* \preceq W_p^*$).

For part two, the result that AI agents are necessarily used as workers when $\zbar\in W$ is intuitive. Indeed, AI agents must be employed either as workers or solvers; otherwise, they would not interact with humans, so humans would match exactly as in the pre-AI equilibrium, preserving pre-AI wages. This would imply $w^*(\zbar)=w(\zbar)>\zbar$, a contradiction. 

Hence, if AI agents are not workers, they must be solvers. However, this would imply that, post-AI, the economy has more solvers and weakly fewer workers than before AI (as $W^* \preceq \{\zbar\}$), violating market clearing. A similar logic explains why AI agents are necessarily used as solvers in equilibrium when $\zbar \in S$. Determining whether AI agents are simultaneously used in all three roles when $\zbar \in W$ or $\zbar \in S$ is not fundamental for our main results. For this reason, we relegate such details to the Online Appendix.

Finally, the third part of the proposition addresses equilibrium prices. The result that $r^*=\zbar$ follows because some compute must be used for independent production, and single-layer automated firms must break even. The wages of the humans employed by $t$A and $b$A firms---those in $W_a^*$ and $S_a^*$---are determined by the zero-profit conditions of these firms, along with the fact that $r^{\post}=\zbar$: \begin{align*} & n(z)[\zbar-w^*(z)]-r^* =0 \Longrightarrow w^{\post}(z)=\zbar(1-1/n(z)), \ \text{for any $z \in W_a^*$} \\
& n(\zbar)[z-r^*]-w^*(z) =0 \Longrightarrow w^*(z)=n(\zbar)(z-\zbar), \ \text{for any $z \in S_a^*$}  \end{align*}

The wages of the human workers and solvers hired by $n$A firms---those in $W_p^*$ and $S_p^*$---are derived following a similar logic as in the pre-AI equilibrium. The only difference is that the most knowledgeable workers and least knowledgeable solvers earn their expected output as independent producers (while they earn strictly more than that in the pre-AI equilibrium). The reason for this is that the equilibrium wage function must be continuous, and there is always some independent production in the post-AI equilibrium (done by AI agents and possibly also humans).

 \section{The Impact of Autonomous AI}\label{sec:effects}

We now analyze the impact of an autonomous AI by comparing the pre- and post-AI equilibrium. Figure \ref{fig:eq1} illustrates this comparison, showing substantial changes in three areas: (i) occupational choice, (ii) the matching of workers and solvers, and (iii) the distribution of labor income. In the Online Appendix, we further examine how these changes influence the distribution of firm size, productivity, and decentralization.

For brevity, we focus on two cases: (i) "basic" AI, which creates AI agents with the knowledge of pre-AI workers ($\zbar \in \mathrm{int} W$), and (ii) "advanced" AI, which creates AI agents with the knowledge of pre-AI solvers ($\zbar \in \mathrm{int} S$). The results for the knife-edge case where AI has the knowledge of both ($\zbar \in W \cap S$) are in the Online Appendix.
 
 \begin{figure}[!b]
    \centering
  \begin{tikzpicture}[xscale=5.8140, yscale=3.42]
 \draw[->] (-.02+0, 0)--(1.03+0,  0);
\draw[->] (0+0, -.02)--(0+0,2.05);
\draw[dotted,thick] (0+0,0)--(1+0,1);
\node[right] at (1.03+0,0) {\Large $z$}; 
\draw[thick, gray, domain=0:{3-sqrt(5)}, samples=200] plot (\x+0, {3-sqrt(5)- (1/2*(3-sqrt(5))*(1+1/4*(3-sqrt(5))))/(1+1/2-1/2*(3-sqrt(5))) *(1-\x)+1/2*\x^2/2});
\draw[thick, gray, domain={3-sqrt(5)}:1, samples=200] plot (\x+0, {(1/2+2*(3-sqrt(5)))/(1+1/2*(1-(3-sqrt(5))))-sqrt(1+2*(3-sqrt(5))*2-2*\x*2)});
\draw[dashed] ({3-sqrt(5)+0},0)--({3-sqrt(5)+0},{3-sqrt(5)- (1/2*(3-sqrt(5))*(1+1/4*(3-sqrt(5))))/(1+1/2-1/2*(3-sqrt(5))) *(1-(3-sqrt(5)))  +1/2*(3-sqrt(5))^2/2}) ;
\draw (1+0,-0.02)--(1+0,0.02); 
\node[below] at (0+0,0) {$0$}; 
\node[below] at (1+0,0) {$1$}; 
\draw [decorate, decoration={brace, amplitude=10pt},gray] ({3-sqrt(5)+0},0)--(0+0,0);
\node[below] at ({(3-sqrt(5))/2+0},-.1) {$\color{gray} W$}; 
\draw [decorate, decoration={brace, amplitude=10pt},gray] (1+0,0) -- ({3-sqrt(5)+0},0);
\node[below] at ({3-sqrt(5)+(1-(3-sqrt(5)))/2+0.01+0},-.1) {$\color{gray} S$}; 
\node[above] at (.75+0,1.4) {$w^{*}(z)$}; 
\draw[ultra thick, domain=0:0.426, samples=100,black] plot (\x+0, {0.26656 + 0.26656*\x + 0.25000*\x^2});
\draw[ultra thick,black] (0.425+0,0.425)--(0.53311+0,0.53311);
\draw[ultra thick, domain=(0.53311-0.001):0.7009, samples=100,black] plot (\x+0, {1.5331 - sqrt(3.1323 - 3.9998*\x)});
\draw[ultra thick, domain=(0.70046-0.001):1, samples=100,black] plot (\x+0, {-1.4783 + 3.4783*\x});
 \draw [decorate, decoration={brace, amplitude=10pt}] (0+0,2.1)--(0.425+0,2.1);
 \node[above] at (0.425*.5+0+.005,2.18) {$W_p^{\post}$}; 
 \draw [decorate, decoration={brace, amplitude=10pt}] (0.425+0,2.1)--(0.53311+0,2.1);
 \node[above] at ({0.425+(0.53311-0.425)/2+0+.015},2.225) {$I^{\post}$}; 
 \draw [decorate, decoration={brace, amplitude=10pt}] (0.53311+0,2.1)--(0.70046+0,2.1);
 \node[above] at ({0.53311+(0.70046-0.53311)/2+0+.0075},2.18) {$S_p^{\post}$}; 
 \draw [decorate, decoration={brace, amplitude=10pt}] (0.70046+0,2.1)--(1+0,2.1);
 \node[above] at ({0.70046+(1-0.70046)/2+0+.0075},2.19) {$S_a^{\post}$}; 
  \node[left] at (0,0.425) {$\zbar$};
\draw[dotted] (.425+0, 0.425)--(.425+0,2.1);
\draw[dotted] (0.53311+0, 0)--(0.53311+0,2.1);
\draw[dotted] (0.70046, 0)--(0.70046+0,1.43);
\draw[dotted] (0.70046, 1.6)--(0.70046+0,2.1);
\draw[dotted] (1+0, 1.68)--(1+0,2.1);
\draw[dotted] (1+0, 1.55)--(1+0,0);
 \node[above] at (0.425+0,0) {$\zbar$};
\draw (0.425+0,0.02)--(0.425+0,-0.02);
\draw[dashed] (0.425+0, .11)--(0.425+0, 0.425);
\draw[dashed] (0.425+0, 0.425)--(0+0, 0.425);
\node[right] at (0+.96,1.63) {$\color{gray} w(z)$}; 
\node[left,gray] at (0,1.5778) {$1.58$}; 
\draw[dotted] (0,1.5778)--(1,1.5778);
\node[left] at (0,2) {$2$}; 
\draw[dotted] (0, 2)--(1,2);
\node[above] at ({1/2},-0.5) {  (a) Basic AI (i.e., $\zbar \in  \mathrm{int} W$) };  
 
 
 \draw[->] (1.3-.02, 0)--(1.3+1.03,  0);
\draw[->] (1.3+0, -.02)--(1.3+0,2.05);
\draw[dotted,thick] (1.3+0,0)--(1.3+1,1);
\node[right] at (1.3+1.03,0) {\Large $z$}; 
\draw[thick, gray, domain=0:{3-sqrt(5)}, samples=200] plot (1.3+\x, {3-sqrt(5)- (1/2*(3-sqrt(5))*(1+1/4*(3-sqrt(5))))/(1+1/2-1/2*(3-sqrt(5))) *(1-\x)+1/2*\x^2/2});
\draw[thick, gray, domain={3-sqrt(5)}:1, samples=200] plot (1.3+\x, {(1/2+2*(3-sqrt(5)))/(1+1/2*(1-(3-sqrt(5))))-sqrt(1+2*(3-sqrt(5))*2-2*\x*2)});
\draw[dashed, gray] ({1.3+3-sqrt(5)},0)--({1.3+3-sqrt(5)},{3-sqrt(5)- (1/2*(3-sqrt(5))*(1+1/4*(3-sqrt(5))))/(1+1/2-1/2*(3-sqrt(5))) *(1-(3-sqrt(5)))  +1/2*(3-sqrt(5))^2/2}) ;
\draw (1.3+1,-0.02)--(1.3+1,0.02); 
\node[below] at (1.3+0,0) {$0$}; 
\node[below] at (1.3+1,0) {$1$}; 
\node[right] at (1.3+.96,1.63) {$\color{gray} w(z)$}; 
\node[above] at (1.3+0.85,0) {$\zbar$};
\node[left] at (1.3,0.85) {$\zbar$};
\draw[dashed] (1.3,0.85)--(1.3+.85,.85);       
\node[left] at (1.3+0.01,0.99) {$\color{gray} w(\zbar)$};
\draw[dashed] (1.3,0.97977)--(1.3+.85,0.97977);       
\draw[dotted] (1.3+1,0)--(1.3+1,1.6); 
\draw[dotted] (1.3+1,1.70)--(1.3+1,2.1); 
\draw[dashed] (1.3+.85,0.15)--(1.3+.85,0.97977);       
\draw[dashed] (1.3+.85,0)--(1.3+.85,0.02);              
\draw [decorate, decoration={brace, amplitude=10pt},gray] ({3-sqrt(5)+1.3},0)--(0+1.3,0);
\node[below] at ({(3-sqrt(5))/2+1.3},-.1) {$\color{gray} W$}; 
\draw [decorate, decoration={brace, amplitude=10pt},gray] (1+1.3,0) -- ({3-sqrt(5)+1.3},0);
\node[below] at ({3-sqrt(5)+(1-(3-sqrt(5)))/2+0.01+1.3},-.1) {$\color{gray} S$}; 
 \draw [decorate, decoration={brace, amplitude=10pt}] (0+1.3,2.1)--(.345025+1.3,2.1);
 \node[above] at (.345025/2+1.3,2.18) {$W_a^{\post}$}; 
 \draw [decorate, decoration={brace, amplitude=10pt}] (.345025+1.3,2.1)--(.85+1.3,2.1);
 \node[above] at ({.345025+(.85-.345025)/2+1.3},2.18) {$W_p^{\post}$}; 
  \draw [decorate, decoration={brace, amplitude=10pt}] (.85+1.3,2.1)--(.948+1.3,2.1);
 \node[above] at ({.86+(.948-.85)/2+1.3-0.02},2.18) {$S_p^*$}; 
   \draw [decorate, decoration={brace, amplitude=10pt}] (.948+1.3,2.1)--(1+1.3,2.1);
 \node[above] at ({.948+(1-.948)/2+1.3+0.01},2.18) {$S_a^*$}; 
\draw[ultra thick, domain=0:.345025, samples=100] plot (\x+1.3, {.425+.425*\x});
\draw[ultra thick, domain=.345025:.85, samples=100] plot (\x+1.3, {.45476+.252488*\x+.25*\x^2});
\draw[ultra thick, domain=.85:.953, samples=100] plot (\x+1.3, {1.50498-sqrt(3.82899-4*\x)});
\draw[ultra thick, domain=.951623:1, samples=100] plot (\x+1.3, {-11.3333+13.3333*\x});
\node[above,black] at (1.3+.18,0.52) {$ w^*(z)$};  
\draw[dotted] (0.345025+1.3, 0)--(0.345025+1.3, 2.1);
\draw[dotted] (.85+1.3, 0.97977)--(.85+1.3, 2.1);
\draw[dotted] (.948+1.3, 0)--(.948+1.3, 2.1);
\node[left,gray] at (0+1.3,1.5778) {$1.58$}; 
\draw[dotted] (0+1.3,1.5778)--(1+1.3,1.5778);
\node[left] at (0+1.3,2) {$2$}; 
\draw[dotted] (0+1.3, 2)--(1+1.3,2);
\node[above] at ({1/2+1.3},-0.5) { (b) Advanced AI (i.e., $\zbar \in \mathrm{int}  S$)};  

 \end{tikzpicture}
\captionsetup{justification=centering}
 \caption{Comparison of the Pre- and Post-AI Equilibrium\\ \justifying 
  \vspace{0.5mm}
 \footnotesize \noindent  \textit{Notes}. Distribution of knowledge: $G(z)=z$. Parameter values: For both panels, $h=1/2$ ($<h_0= 3/4$). For panel (a), $\zbar=0.425$, while for panel (b), $\zbar=0.85$. In the post-AI equilibrium depicted in panel (a), AI is used as a worker and independent producer. In panel (b), AI is used in all three possible roles.} \label{fig:eq1}
\end{figure}

\subsection{Occupational Choice and Worker-Solver Matching} \label{sec:displacement}

We begin by analyzing the effects of AI on occupational choice. 

\begin{prop} \label{prop:displacement} If $\zbar \in \mathrm{int} W$, AI displaces humans from routine knowledge work to specialized problem solving, i.e., $W^* \subset W$ and $S^* \supset S$. In contrast, if $\zbar \in \mathrm{int} S$, AI displaces humans in the opposite direction, i.e., $W^* \supset W$ and $S^* \subset S$. \end{prop}

Intuitively, when AI is basic ($\zbar \in \mathrm{int} W$), it serves as a relatively inexpensive technology for routine work, lowering workers' wages.  This increases demand for solvers to match with the less expensive, more abundant workers (both humans and AI), drawing marginal pre-AI workers into specialized problem solving. Conversely, when AI is advanced ($\zbar \in \mathrm{int} S$), it serves as a relatively inexpensive technology for specialized problem solving. This increases the demand for workers to match with the less expensive, more abundant solvers (both humans and AI), inducing the marginal pre-AI solvers to shift to routine knowledge work.

Next, we examine how AI affects the productivity of pre-AI workers who continue as workers and the span of control of pre-AI solvers who continue as solvers. We define a worker's (average) productivity as her expected output per unit of time. Thus, it is equal to the knowledge of the solver with whom she is matched. Similarly, a solver's span of control is equal to the number of workers (human or AI) under her supervision. Hence, a solver's span of control is increasing in the knowledge of the workers with whom she is matched. For the following result, recall that $e(z)$ is the pre-AI equilibrium employee function:

\begin{prop} \label{prop:matches} $\quad$ \begin{itemize}[leftmargin=*]	
\item If $\zbar \in \mathrm{int} W$, then the productivity of $z \in W^* \subset W$ is strictly lower post-AI than pre-AI. Moreover, the span of control of $z \in S \subset S^*$ is strictly larger post-AI than pre-AI if $e(z)<\zbar$, and strictly smaller post-AI than pre-AI if $e(z)>\zbar$. 
\item If $\zbar \in \mathrm{int} S$, then the productivity of $z \in W \subset W^*$ is strictly higher post-AI than pre-AI if $z < e(\zbar)$, and strictly lower post-AI than pre-AI if $z> e(\zbar)$. Moreover, the span of control of $z \in S^* \subset S$ is strictly larger post-AI than pre-AI. \end{itemize}  \end{prop}

Proposition \ref{prop:matches} is illustrated in Figure \ref{fig:intuition} when $\zbar \in \mathrm{int} W$ and AI is used as a worker and as an independent producer (but not as a solver). As the figure shows, AI worsens the matches of all pre-AI workers who remain workers (i.e., those humans in $W_p^*$), reducing their productivity. Moreover, AI improves the match of the least knowledgeable solvers who are solvers both pre- and post-AI---increasing their span of control---but worsens the match of the most knowledgeable solvers.

  \begin{figure}[t!]
    \centering
 \begin{tikzpicture}[xscale=5.8140, yscale=3.42]
 
\draw[->] (-.02+1.2, 0+0)--(1.03+1.2,  0+0);
\node[right] at (1.03+1.2,0+0) {\Large $z$}; 
\draw (1+1.2,-0.02+0)--(1+1.2,0.02+0); 
\node[below] at (0+1.2,0+0) {$0$}; 
\node[below] at (1+1.2,0+0) {$1$}; 
\draw (0.425+1.2,-0.02+0)--(0.425+1.2,0.02+0); 
\draw[dotted] (1+1.2, 0+0)--(1+1.2,0.35+0.01+0);
\draw [decorate, decoration={brace, amplitude=10pt, aspect=2/3},gray] ({3-sqrt(5)+1.2},0+0)--(0+1.2,0+0);
\node[below] at ({(3-sqrt(5))/3+1.2},-.1+0) {$W$}; 
\draw [decorate, decoration={brace, amplitude=10pt},gray] (1+1.2,0+0) -- ({3-sqrt(5)+1.2},0+0);
\node[below] at ({3-sqrt(5)+(1-(3-sqrt(5)))/2+0.01+1.2},-.1+0) {$ S$}; 
\node[below] at (0.425+1.2,-.04+0) {$\zbar$};
 \draw [decorate, decoration={brace, amplitude=10pt}] (0+1.2,0.35+0.01+0)--(0.425+1.2,0.35+0.01+0);
 \node[above] at (0.425/2+1.2+.005,0.43+0.01+0) {$W_p^{\post}$}; 
 \draw [decorate, decoration={brace, amplitude=10pt}] (0.425+1.2,0.35+0.01+0)--(0.53311+1.2,0.35+0.01+0);
 \node[above] at ({0.425+(0.53311-0.425)/2+1.2+.015},0.43+0.01+0.035+0) {$I^{\post}$}; 
  \draw [decorate, decoration={brace, amplitude=10pt}] (0.53311+1.2,0.35+0.01+0)--(0.70046+1.2,0.35+0.01+0);
 \node[above] at ({0.53311+(0.70046-0.53311)/2+1.2+.0075},0.43+0.01+0) {$S_p^{\post}$}; 
   \draw [decorate, decoration={brace, amplitude=10pt}] (0.70046+1.2,0.35+0.01+0)--(1+1.2,0.35+0.01+0);
 \node[above] at ({0.70046+(1-0.70046)/2+1.2},0.43+0.01+0) {$S_a^{\post}$}; 
\draw[dotted] (0+1.2, 0+0)--(0+1.2, 0.35+0.01+0);
\draw[dotted] (0.425+1.2, 0+0)--(0.425+1.2, 0.35+0.01+0);
\draw[dotted] (0.53311+1.2, 0+0)--(0.53311+1.2, 0.35+0.01+0);
\draw[dotted] (0.70046+1.2, 0+0)--(0.70046+1.2, 0.35+0.01+0);
\draw[thick, densely dashed, <->, out=60, in=120] (0+1.2,0) to (0.7639+1.2,0);   
\draw[thick, densely dashed, <->, out=60, in=120] (0.425+1.2,0) to (0.931275773+1.2,0);   
\node[above] at ({1/2+1.2},-0.45+0) { (a) Pre-AI Matches of $z \in W^* \subset W$ };   
 
\draw[->] (-.02+2.5, 0+0)--(1.03+2.5,  0+0);
\node[right] at (1.03+2.5,0+0) {\Large $z$}; 
\draw (1+2.5,-0.02+0)--(1+2.5,0.02+0); 
\node[below] at (0+2.5,0+0) {$0$}; 
\node[below] at (1+2.5,0+0) {$1$}; 
\draw (0.425+2.5,-0.02+0)--(0.425+2.5,0.02+0); 
\draw[dotted] (1+2.5, 0+0)--(1+2.5,0.35+0.01+0);
\draw [decorate, decoration={brace, amplitude=10pt, aspect=2/3},gray] ({3-sqrt(5)+2.5},0+0)--(0+2.5,0+0);
\node[below] at ({(3-sqrt(5))/3+2.5},-.1+0) {$W$}; 
\draw [decorate, decoration={brace, amplitude=10pt},gray] (1+2.5,0+0) -- ({3-sqrt(5)+2.5},0+0);
\node[below] at ({3-sqrt(5)+(1-(3-sqrt(5)))/2+0.01+2.5},-.1+0) {$ S$}; 
\node[below] at (0.425+2.5,-.04+0) {$\zbar$};
 \draw [decorate, decoration={brace, amplitude=10pt}] (0+2.5,0.35+0.01+0)--(0.425+2.5,0.35+0.01+0);
 \node[above] at (0.425/2+2.5+.005,0.43+0.01+0) {$W_p^{\post}$}; 
 \draw [decorate, decoration={brace, amplitude=10pt}] (0.425+2.5,0.35+0.01+0)--(0.53311+2.5,0.35+0.01+0);
 \node[above] at ({0.425+(0.53311-0.425)/2+2.5+.015},0.43+0.01+0.035+0) {$I^{\post}$}; 
  \draw [decorate, decoration={brace, amplitude=10pt}] (0.53311+2.5,0.35+0.01+0)--(0.70046+2.5,0.35+0.01+0);
 \node[above] at ({0.53311+(0.70046-0.53311)/2+2.5+.0075},0.43+0.01+0) {$S_p^{\post}$}; 
   \draw [decorate, decoration={brace, amplitude=10pt}] (0.70046+2.5,0.35+0.01+0)--(1+2.5,0.35+0.01+0);
 \node[above] at ({0.70046+(1-0.70046)/2+2.5},0.43+0.01+0) {$S_a^{\post}$}; 
\draw[dotted] (0+2.5, 0+0)--(0+2.5, 0.35+0.01+0);
\draw[dotted] (0.425+2.5, 0+0)--(0.425+2.5, 0.35+0.01+0);
\draw[dotted] (0.53311+2.5, 0+0)--(0.53311+2.5, 0.35+0.01+0);
\draw[dotted] (0.70046+2.5, 0+0)--(0.70046+2.5, 0.35+0.01+0);
\draw[thick, densely dashed, <->, out=60, in=120] (0+2.5,0) to (0.7639+2.5,0);   
\draw[thick, densely dashed, <->, out=60, in=120] (0.7639+2.5,0) to (1+2.5,0);   
\node[above] at ({1/2+2.5},-0.45+0) { (b) Pre-AI Matches of $z \in S \subset S^*$ };   


\draw[->] (-.02+1.2, 0-1.175)--(1.03+1.2,  0-1.175);
\node[right] at (1.03+1.2,0-1.175) {\Large $z$}; 
\draw (1+1.2,-0.02-1.175)--(1+1.2,0.02-1.175); 
\node[below] at (0+1.2,0-1.175) {$0$}; 
\node[below] at (1+1.2,0-1.175) {$1$}; 
\draw (0.425+1.2,-0.02-1.175)--(0.425+1.2,0.02-1.175); 
\draw[dotted] (1+1.2, 0-1.175)--(1+1.2,0.35+0.01-1.175);
\draw [decorate, decoration={brace, amplitude=10pt, aspect=2/3},gray] ({3-sqrt(5)+1.2},0-1.175)--(0+1.2,0-1.175);
\node[below] at ({(3-sqrt(5))/3+1.2},-.1-1.175) {$W$}; 
\draw [decorate, decoration={brace, amplitude=10pt},gray] (1+1.2,0-1.175) -- ({3-sqrt(5)+1.2},0-1.175);
\node[below] at ({3-sqrt(5)+(1-(3-sqrt(5)))/2+0.01+1.2},-.1-1.175) {$ S$}; 
\node[below] at (0.425+1.2,-.04-1.175) {$\zbar$};
 \draw [decorate, decoration={brace, amplitude=10pt}] (0+1.2,0.35+0.01-1.175)--(0.425+1.2,0.35+0.01-1.175);
 \node[above] at (0.425/2+1.2+.005,0.43+0.01+0-1.175) {$W_p^{\post}$}; 
 \draw [decorate, decoration={brace, amplitude=10pt}] (0.425+1.2,0.35+0.01-1.175)--(0.53311+1.2,0.35+0.01-1.175);
 \node[above] at ({0.425+(0.53311-0.425)/2+1.2+.015},0.43+0.01+0.035+0-1.175) {$I^{\post}$}; 
  \draw [decorate, decoration={brace, amplitude=10pt}] (0.53311+1.2,0.35+0.01-1.175)--(0.70046+1.2,0.35+0.01-1.175);
 \node[above] at ({0.53311+(0.70046-0.53311)/2+1.2+.0075},0.43+0.01+0-1.175) {$S_p^{\post}$}; 
   \draw [decorate, decoration={brace, amplitude=10pt}] (0.70046+1.2,0.35+0.01-1.175)--(1+1.2,0.35+0.01-1.175);
 \node[above] at ({0.70046+(1-0.70046)/2+1.2},0.43+0.01+0-1.175) {$S_a^{\post}$}; 
\draw[dotted] (0+1.2, 0-1.175)--(0+1.2, 0.35+0.01-1.175);
\draw[dotted] (0.425+1.2, 0-1.175)--(0.425+1.2, 0.35+0.01-1.175);
\draw[dotted] (0.53311+1.2, 0-1.175)--(0.53311+1.2, 0.35+0.01-1.175);
\draw[dotted] (0.70046+1.2, 0-1.175)--(0.70046+1.2, 0.35+0.01-1.175);
\draw[thick, densely dashed, <->, out=60, in=120] (0+1.2,0-1.175) to (0.53311+1.2,0-1.175);   
\draw[thick, densely dashed, <->, out=60, in=120] (0.425+1.2,0-1.175) to (0.70046+1.2,0-1.175);   
\node[above] at ({1/2+1.2},-0.45+0-1.175) { (c) Post-AI Matches of $z \in W^* \subset W$ };   
 

\draw[->] (-.02+2.5, 0-1.175)--(1.03+2.5,  0-1.175);
\node[right] at (1.03+2.5,0-1.175) {\Large $z$}; 
\draw (1+2.5,-0.02-1.175)--(1+2.5,0.02-1.175); 
\node[below] at (0+2.5,0-1.175) {$0$}; 
\node[below] at (1+2.5,0-1.175) {$1$}; 
\draw (0.425+2.5,-0.02-1.175)--(0.425+2.5,0.02-1.175); 
\draw[dotted] (1+2.5, 0-1.175)--(1+2.5,0.35+0.01-1.175);
\draw [decorate, decoration={brace, amplitude=10pt, aspect=2/3},gray] ({3-sqrt(5)+2.5},0-1.175)--(0+2.5,0-1.175);
\node[below] at ({(3-sqrt(5))/3+2.5},-.1-1.175) {$W$}; 
\draw [decorate, decoration={brace, amplitude=10pt},gray] (1+2.5,0-1.175) -- ({3-sqrt(5)+2.5},0-1.175);
\node[below] at ({3-sqrt(5)+(1-(3-sqrt(5)))/2+0.01+2.5},-.1-1.175) {$ S$}; 
\node[below] at (0.425+2.5,-.04-1.175) {$\zbar$};
 \draw [decorate, decoration={brace, amplitude=10pt}] (0+2.5,0.35+0.01-1.175)--(0.425+2.5,0.35+0.01-1.175);
 \node[above] at (0.425/2+2.5+.005,0.43+0.01+0-1.175) {$W_p^{\post}$}; 
 \draw [decorate, decoration={brace, amplitude=10pt}] (0.425+2.5,0.35+0.01-1.175)--(0.53311+2.5,0.35+0.01-1.175);
 \node[above] at ({0.425+(0.53311-0.425)/2+2.5+.015},0.43+0.01+0.035+0-1.175) {$I^{\post}$}; 
  \draw [decorate, decoration={brace, amplitude=10pt}] (0.53311+2.5,0.35+0.01-1.175)--(0.70046+2.5,0.35+0.01-1.175);
 \node[above] at ({0.53311+(0.70046-0.53311)/2+2.5+.0075},0.43+0.01+0-1.175) {$S_p^{\post}$}; 
   \draw [decorate, decoration={brace, amplitude=10pt}] (0.70046+2.5,0.35+0.01-1.175)--(1+2.5,0.35+0.01-1.175);
 \node[above] at ({0.70046+(1-0.70046)/2+2.5},0.43+0.01+0-1.175) {$S_a^{\post}$}; 
\draw[dotted] (0+2.5, 0-1.175)--(0+2.5, 0.35+0.01-1.175);
\draw[dotted] (0.425+2.5, 0-1.175)--(0.425+2.5, 0.35+0.01-1.175);
\draw[dotted] (0.53311+2.5, 0-1.175)--(0.53311+2.5, 0.35+0.01-1.175);
\draw[dotted] (0.70046+2.5, 0-1.175)--(0.70046+2.5, 0.35+0.01-1.175);
\draw[thick, densely dashed, <->, out=60, in=120] (0.425+2.5,0-1.175) to (0.7639+2.5,0-1.175);   
\draw[thick, densely dashed, <->, out=60, in=120] (0.425+2.5,0-1.175) to (1+2.5,0-1.175);   
\node[above] at ({1/2+2.5},-0.45-1.175) { (d) Post-AI Matches of $z \in S \subset S^*$ };   
  \end{tikzpicture}
\captionsetup{justification=centering}
 \caption{An Illustration of Proposition \ref{prop:matches} - $\zbar \in \mathrm{int} W$ \\ \justifying 
 \vspace{0.5mm}
 \footnotesize \noindent  \textit{Notes}. Distribution of knowledge: $G(z)=z$. Parameter values: $h=1/2$ and $\zbar=0.425$. In the post-AI equilibrium shown in the figure, AI is used as a worker and independent producer. The dashed arrows illustrate the matching between workers and solvers.} \label{fig:intuition}
\end{figure}

For intuition, consider the case $\zbar \in \mathrm{int} W$ (the argument for $\zbar \in \mathrm{int} S$ is similar). In this case, (i) AI agents are the best workers in the post-AI economy and are therefore assisted by the most knowledgeable solvers (Proposition \ref{prop:eqAI}), and (ii) the knowledge of the marginal solvers decreases with AI's introduction (Proposition \ref{prop:displacement}). Both effects worsen the pool of solvers available for those pre-AI workers who remain workers post-AI, thus reducing their productivity. 

Regarding the solvers who are not occupationally displaced, the match of the most knowledgeable solvers worsens with AI's introduction because post-AI, they assist production by AI, while pre-AI, they assist humans who are more knowledgeable than AI. However, the match of the least knowledgeable solvers improves, as post-AI, the least knowledgeable workers are assisted by the newly appointed solvers (humans and possibly AI), who are less knowledgeable than the least knowledgeable pre-AI solvers. 

There is anecdotal evidence of these reorganizations in professional services. For example, a 2022 survey of M\&A lawyers found that AI, by automating routine knowledge work such as document review, has allowed junior lawyers to focus on more complex tasks, including client engagement and legal analysis \citep{litera}. This aligns with Proposition \ref{prop:displacement}, as the introduction of basic AI pushes marginal human workers into problem-solving roles.

Similarly, in investment banking, \cite{beane2024inverted} document that junior analysts are becoming increasingly distanced from senior partners as AI takes over tasks that once required junior involvement. This is consistent with Proposition \ref{prop:matches}, which indicates that basic AI forces those workers who remain in their positions to rely on less knowledgeable individuals for support, as the most knowledgeable pre-AI solvers shift their focus to assisting AI-driven production.

 \subsection{Labor Income} \label{sec:wages}
 
We now analyze the effects of AI on total labor income and its distribution. To begin, note that total labor income increases with AI's introduction: if this were not the case, then the set of firms that existed pre-AI could collectively earn strictly positive aggregate profits post-AI—a contradiction. This also immediately implies that total output increases with the introduction of AI, as it is the sum of labor income and capital income, and pre-AI, capital income is zero in this knowledge economy.

Next, we examine AI’s impact on the distribution of labor income. Analyzing these distributional effects is challenging due to the presence of two potentially opposing forces. On one hand, AI alters firm composition and, therefore, the quality of matches (as shown by Proposition \ref{prop:matches}). On the other hand, by automating humans with knowledge \( \zbar \), AI shifts the relative scarcities of different knowledge levels, influencing how each firm’s output is divided between workers and solvers. 

To begin, note that the individual with knowledge $\zbar$ always loses from AI, i.e., $w^{\post}(\zbar)<w(\zbar)$. Define then $B$ and $T$ as the sets of individuals below and above $\zbar$ who benefit from AI, respectively, i.e., $B \equiv \{ z \in  [0,\zbar]: w^*(z)>w(z) \}$ and $ T \equiv  \{ z \in  [\zbar,1]: w^*(z)>w(z) \}$. As we show in the Online Appendix, \( B \) takes the form \( [0, z_b) \), while \( T \) takes the form \( (z_t,1] \) for some \( z_b \geq 0 \) and \( z_t \leq 1 \). In other words, the individuals who gain from AI are necessarily located at the extremes of the knowledge distribution. The next proposition characterizes when \( B \) and \( T \) are non-empty:

 \begin{prop} \label{prop:wages} There are winners at the bottom if AI is good enough, i.e., $B \neq \emptyset$ if and only if $\zbar>\Zeta$, where $\Zeta \in \mathrm{int} W$. In contrast, there are always winners at the top, i.e., $T \neq \emptyset$ for all $\zbar \in [0,1)$. \end{prop}  
  
Figure \ref{fig:eq1}, at the beginning of this section, provides an illustration of Proposition \ref{prop:wages}. As shown in panel (a) of the figure, only the most knowledgeable humans benefit when $\zbar $ is relatively low. In contrast, as shown in panel (b), both the least and the most knowledgeable humans benefit when $\zbar$ is relatively high.

We now provide intuition for this proposition. We do this using Figure \ref{fig:wages}, which builds on Figure \ref{fig:eq1}(a) and illustrates the pre- and post-AI scenarios when $\zbar \in \mathrm{int} W$. Since \( B \) takes the form \([0, z_b)\) and \( T \) takes the form \((z_t, 1]\), to verify the existence of winners below or above \( \zbar \), it suffices to analyze AI's impact on the wages of the least and most knowledgeable individuals. Accordingly, Figure \ref{fig:wages}(a) shows how AI affects the match of individuals with $z = 0$ and the wages of their match before and after AI, while Figure \ref{fig:wages}(b) provides the same analysis for individuals with $z = 1$.

\begin{figure}[t!]
     \centering
  \begin{tikzpicture}[xscale=5.8140, yscale=3.42]
\draw[->] (-.02+1.2, 0)--(1.03+1.2,  0);
\draw[->] (0+1.2, -.02)--(0+1.2,2.05);
\draw[dotted,thick] (0.11+1.2,0.11)--(1+1.2,1);
\node[right] at (1.03+1.2,0) {\Large $z$}; 
\draw[thick, gray, domain=0:{3-sqrt(5)}, samples=200] plot (\x+1.2, {3-sqrt(5)- (1/2*(3-sqrt(5))*(1+1/4*(3-sqrt(5))))/(1+1/2-1/2*(3-sqrt(5))) *(1-\x)  
+1/2*\x^2/2});
\draw[thick, gray, domain={3-sqrt(5)}:1, samples=200] plot (\x+1.2, {(1/2+2*(3-sqrt(5)))/(1+1/2*(1-(3-sqrt(5))))-sqrt(1+2*(3-sqrt(5))*2-2*\x*2)});
\draw[dashed, gray] ({3-sqrt(5)+1.2},0)--({3-sqrt(5)+1.2},{3-sqrt(5)- (1/2*(3-sqrt(5))*(1+1/4*(3-sqrt(5))))/(1+1/2-1/2*(3-sqrt(5))) *(1-(3-sqrt(5)))  +1/2*(3-sqrt(5))^2/2}) ;
\draw (1+1.2,-0.02)--(1+1.2,0.02); 
\node[below] at (0+1.2,0) {$0$}; 
\node[below] at (1+1.2,0) {$1$}; 
\node[above] at (1+1.2,1.55) {$\color{gray} w(z)$};

\draw[ultra thick, domain=0:0.426, samples=100,black] plot (\x+1.2+0, {0.26656 + 0.26656*\x + 0.25000*\x^2});
\draw[ultra thick,black] (0.425+0+1.2,0.425)--(0.53311+0+1.2,0.53311);
\draw[ultra thick, domain=(0.53311-0.001):0.7009, samples=100,black] plot (\x+1.2, {1.5331 - sqrt(3.1323 - 3.9998*\x)});
\draw[ultra thick, domain=(0.70046-0.001):1, samples=100,black] plot (\x+1.2, {-1.4783 + 3.4783*\x});

 \draw [decorate, decoration={brace, amplitude=10pt}] (0+1.2,2.1)--(0.425+1.2,2.1);
 \node[above] at (0.425*.5+1.2+.005,2.18) {$W_p^{\post}$}; 
 \draw [decorate, decoration={brace, amplitude=10pt}] (0.425+1.2,2.1)--(0.53311+1.2,2.1);
 \node[above] at ({0.425+(0.53311-0.425)/2+1.2+.015},2.225) {$I^{\post}$}; 
 \draw [decorate, decoration={brace, amplitude=10pt}] (0.53311+1.2,2.1)--(0.70046+1.2,2.1);
 \node[above] at ({0.53311+(0.70046-0.53311)/2+1.2+.0075},2.18) {$S_p^{\post}$}; 
 \draw [decorate, decoration={brace, amplitude=10pt}] (0.70046+1.2,2.1)--(1+1.2,2.1);
 \node[above] at ({0.70046+(1-0.70046)/2+1.2+.0075},2.19) {$S_a^{\post}$}; 
\draw[dashed] (0.425+1.2, 0)--(0.425+1.2, 0.425); 
\draw[dotted] (1+1.2, 0)--(1+1.2,1.55);
\draw[dotted] (1+1.2, 1.68)--(1+1.2,2.1);
\draw [decorate, decoration={brace, amplitude=10pt, aspect=2/3},gray] ({3-sqrt(5)+1.2},0)--(0+1.2,0);
\node[below] at ({(3-sqrt(5))/3+1.2},-.1) {$\color{gray} W$}; 
\draw [decorate, decoration={brace, amplitude=10pt},gray] (1+1.2,0) -- ({3-sqrt(5)+1.2},0);
\node[below] at ({3-sqrt(5)+(1-(3-sqrt(5)))/2+0.01+1.2},-.1) {$\color{gray} S$}; 
 \draw[dashed] (0.425+1.2, 0)--(0.425+1.2, 0.57512); 
 \node[left] at (0+1.2,0.425) {$\zbar$};
 \draw[dashed] (0+1.2, 0.57512)--(0.425+1.2, 0.57512); 
 \node[left] at (0.01+1.2,0.59) {$\color{gray} w(\zbar)$};
\draw[dashed] (1.2+0, 0.425)--(1.2+0.425, 0.425); 
\node[below] at (0.425+1.2,-.04) {$\zbar$};
\node[above,black] at (.8+1.2,1.5777) {$w^*(z)$}; 
 \draw[dotted] (.425+1.2, 0.57512)--(.425+1.2, 2.1);
\draw[dotted] (0.53311+1.2, 0)--(0.53311+1.2, 2.1);
\draw[dotted] (0.70046+1.2, 0)--(0.70046+1.2, 2.1);
\node[left,gray] at (0+1.2,1.5778) {$1.58$}; 
\draw[dotted] (0+1.2,1.5778)--(1+1.2,1.5778);
\node[left] at (0+1.2,2) {$2$}; 
\draw[dotted] (0+1.2, 2)--(1+1.2,2);

 \node[mark size=3pt] at (0.53311+1.2,0.53311) {\pgfuseplotmark{*}}; 
  \node[mark size=3pt,gray] at (0.7639+1.2,0.81373) {\pgfuseplotmark{*}}; 
\draw[thick, densely dashed, ->, out=60, in=120, gray] (0+1.2,0) to (.76+1.2,0);
 \draw[thick, densely dashed, ->, out=60, in=120] (0+1.2,0) to (0.53311+1.2,0);
\node[above] at ({1/2+1.2},-0.5) {  (a) The Least Knowledgeable Individuals };  
 
 
\draw[->] (-.02+2.5, 0)--(1.03+2.5,  0);
\draw[->] (0+2.5, -.02)--(0+2.5,2.05);
\draw[dotted,thick] (0+2.5,0)--(1+2.5,1);
\node[right] at (1.03+2.5,0) {\Large $z$}; 
\draw[thick, gray, domain=0:{3-sqrt(5)}, samples=200] plot (\x+2.5, {3-sqrt(5)- (1/2*(3-sqrt(5))*(1+1/4*(3-sqrt(5))))/(1+1/2-1/2*(3-sqrt(5))) *(1-\x)  
+1/2*\x^2/2});
\draw[thick, gray, domain={3-sqrt(5)}:1, samples=200] plot (\x+2.5, {(1/2+2*(3-sqrt(5)))/(1+1/2*(1-(3-sqrt(5))))-sqrt(1+2*(3-sqrt(5))*2-2*\x*2)});
\draw[dashed, gray] ({3-sqrt(5)+2.5},0)--({3-sqrt(5)+2.5},{3-sqrt(5)- (1/2*(3-sqrt(5))*(1+1/4*(3-sqrt(5))))/(1+1/2-1/2*(3-sqrt(5))) *(1-(3-sqrt(5)))  +1/2*(3-sqrt(5))^2/2}) ;
\draw (1+2.5,-0.02)--(1+2.5,0.02); 
\node[below] at (0+2.5,0) {$0$}; 
\node[below] at (1+2.5,0) {$1$}; 
\node[above] at (1+2.5,1.55) {$\color{gray} w(z)$};

\draw[ultra thick, domain=0:0.426, samples=100,black] plot (\x+2.5+0, {0.26656 + 0.26656*\x + 0.25000*\x^2});
\draw[ultra thick,black] (0.425+0+2.5,0.425)--(0.53311+0+2.5,0.53311);
\draw[ultra thick, domain=(0.53311-0.001):0.7009, samples=100,black] plot (\x+2.5, {1.5331 - sqrt(3.1323 - 3.9998*\x)});
\draw[ultra thick, domain=(0.70046-0.001):1, samples=100,black] plot (\x+2.5, {-1.4783 + 3.4783*\x});

 \draw [decorate, decoration={brace, amplitude=10pt}] (0+2.5,2.1)--(0.425+2.5,2.1);
 \node[above] at (0.425*.5+2.5+.005,2.18) {$W_p^{\post}$}; 
 \draw [decorate, decoration={brace, amplitude=10pt}] (0.425+2.5,2.1)--(0.53311+2.5,2.1);
 \node[above] at ({0.425+(0.53311-0.425)/2+2.5+.015},2.225) {$I^{\post}$}; 
 \draw [decorate, decoration={brace, amplitude=10pt}] (0.53311+2.5,2.1)--(0.70046+2.5,2.1);
 \node[above] at ({0.53311+(0.70046-0.53311)/2+2.5+.0075},2.18) {$S_p^{\post}$}; 
 \draw [decorate, decoration={brace, amplitude=10pt}] (0.70046+2.5,2.1)--(1+2.5,2.1);
 \node[above] at ({0.70046+(1-0.70046)/2+2.5+.0075},2.19) {$S_a^{\post}$}; 
\draw[dashed] (0.425+2.5, 0)--(0.425+2.5, 0.425); 
\draw[dotted] (1+2.5, 0)--(1+2.5,1.55);
\draw[dotted] (1+2.5, 1.68)--(1+2.5,2.1);
\draw [decorate, decoration={brace, amplitude=10pt, aspect=2/3},gray] ({3-sqrt(5)+2.5},0)--(0+2.5,0);
\node[below] at ({(3-sqrt(5))/3+2.5},-.1) {$\color{gray} W$}; 
\draw [decorate, decoration={brace, amplitude=10pt},gray] (1+2.5,0) -- ({3-sqrt(5)+2.5},0);
\node[below] at ({3-sqrt(5)+(1-(3-sqrt(5)))/2+0.01+2.5},-.1) {$\color{gray} S$}; 
 \draw[dashed] (0.425+2.5, 0)--(0.425+2.5, 0.57512); 
 \node[left] at (0+2.5,0.425) {$\zbar$};
 \draw[dashed] (0+2.5, 0.57512)--(0.425+2.5, 0.57512); 
 \node[left] at (0.01+2.5,0.59) {$\color{gray} w(\zbar)$};
\draw[dashed] (2.5+0, 0.425)--(2.5+0.425, 0.425); 
\node[below] at (0.425+2.5,-.04) {$\zbar$};
\node[above,black] at (.8+2.5,1.5777) {$w^*(z)$}; 
 \draw[dotted] (.425+2.5, 0.57512)--(.425+2.5, 2.1);
\draw[dotted] (0.53311+2.5, 0)--(0.53311+2.5, 2.1);
\draw[dotted] (0.70046+2.5, 0)--(0.70046+2.5, 2.1);
\node[left,gray] at (0+2.5,1.5778) {$1.58$}; 
\draw[dotted] (0+2.5,1.5778)--(1+2.5,1.5778);
\node[left] at (0+2.5,2) {$2$}; 
\draw[dotted] (0+2.5, 2)--(1+2.5,2);

 \node[mark size=3pt]  at (0.425+2.5,0.425) {\pgfuseplotmark{*}}; 
   \node[mark size=3pt,gray] at (0.7639+2.5,0.81373) {\pgfuseplotmark{*}}; 
\draw[thick, densely dashed, <-, out=60, in=120,gray] (.76+2.5,0) to (1+2.5,0);
\draw[thick, densely dashed, <-, out=60, in=120,] (0.425+2.5,0) to (1+2.5,0);
\node[above] at ({1/2+2.5},-0.5) {  (b) The Most Knowledgeable Individuals };  

  \end{tikzpicture}
\captionsetup{justification=centering}
 \caption{Decomposing the Effects of AI on Wages \\ \justifying 
  \vspace{0.5mm}
 \footnotesize \noindent  \textit{Notes}. Distribution of knowledge: $G(z)=z$. Parameter values: $h=1/2$ ($<h_0= 3/4$) and $\zbar=0.425$, so AI is used as a worker and independent producer. Panel (a) shows how AI affects the match of $z = 0$ and the wages of their match before and after AI, while panel (b) provides the same analysis for $z = 1$.} \label{fig:wages}
\end{figure}

Consider Figure \ref{fig:wages}(a) first. AI's introduction has two opposing effects on the wages of the least knowledgeable individuals. On the one hand, AI reduces their wages by lowering their productivity, as they are now matched with a less knowledgeable solver (a negative ``match'' effect). On the other hand, AI increases their wages by allowing them to appropriate a larger share of the firm's output (a positive ``share'' effect). This occurs because the least knowledgeable solvers (with whom they are matched) now earn their expected output as independent producers---their wages fall on the 45° line---whereas pre-AI, they earned strictly more. 

Thus, when $\zbar \in \mathrm{int} W$, AI benefits the least knowledgeable individuals if the positive share effect outweighs the negative match effect. This happens when $\zbar$ is sufficiently high (i.e., when $\zbar > \Zeta$, where $\Zeta \in \mathrm{int} W$). Conversely, when $\zbar \in \mathrm{int} S$, the least knowledgeable individuals always benefit from AI’s introduction because both the match and share effects are positive: AI always improves the match of the least knowledgeable individuals in this case (see Proposition \ref{prop:matches}).

Now consider Figure \ref{fig:wages}(b). As in the case of the least knowledgeable individuals,  AI generates both a negative match effect and a positive share effect for the most knowledgeable individuals. However, in this case, the positive share effect always dominates, explaining why there is always a set of winners above $\zbar$. The key driver behind this result is that $h<h_0$, meaning the size of the team each solver assists is large. Consequently, even a small decrease in the wages of the most knowledgeable workers (with whom the most knowledgeable individuals are matched) is amplified by team size, helping the positive share effect outweigh any negative match effect. In a previous version of this paper \citep{idetalamasAI}, we show that AI can sometimes harm the most knowledgeable individuals when $h \ge h_0$.\footnote{The finding that the most knowledgeable individuals always benefit from AI’s introduction depends not only on \( h < h_0 \) but also on the assumption that \( \zbar \in [0,1) \), meaning AI is not “superintelligent.” As we show in the Online Appendix, when \(\zbar = 1 \), the most knowledgeable humans necessarily lose from AI’s introduction.}

\section{Non-Autonomous AI} \label{sec:nonautonomy}

In our baseline model, AI is a technology that transforms units of compute into AI agents that can do exactly the same as humans with knowledge $\zbar$. This type of AI is “autonomous” because it creates agents that function both as co-workers (pursuing production opportunities) and as co-pilots (providing advice). We focus on this case first, as it generates the most anxiety and debate, driven by technology firms developing and deploying AI agents capable of independent cognitive work.

However, for technical or regulatory reasons, many existing AI systems lack the ability to operate autonomously. In this section, we analyze the impact of non-autonomous AI—a technology that creates agents that operate only as co-pilots. Our main finding is that non-autonomous AI tends to benefit the least knowledgeable individuals the most, whereas autonomous AI primarily benefits the most knowledgeable individuals. However, autonomous AI generates higher overall output than non-autonomous AI.

\begin{figure}[!b]
\centering
\includegraphics[scale=0.6]{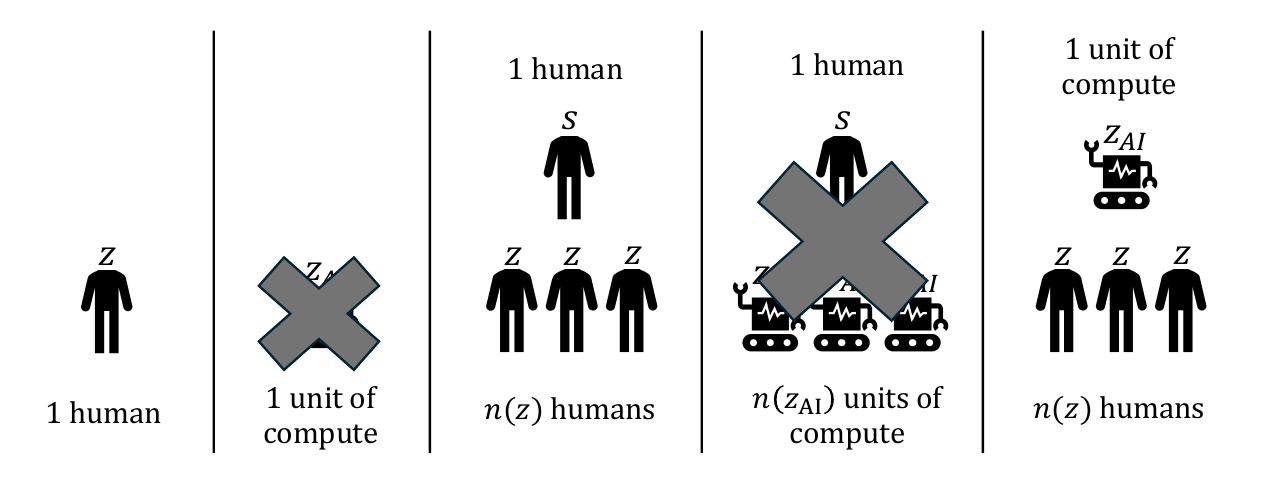}
\caption{The Three Possible Firm Configurations with Non-Autonomous AI} 
\label{fig:jungle3}
\end{figure}

Formally, the model is the same as in Section \ref{sec:setting}, except that AI agents cannot pursue production opportunities. As a result, single-layer and bottom-automated firms do not emerge in equilibrium since they produce no output (see Figure \ref{fig:jungle3}). Thus, \( S_a = \emptyset \) and \( \mu_w = 0 \). In this context, we interpret the amount of compute rented for independent production, $\mu_i$, as the compute that remains idle (i.e., not productive). 

To ensure a meaningful comparison of equilibrium outcomes between autonomous and non-autonomous AI, we maintain the assumption that firms have at most two layers. Additionally, we assume that compute availability is identical and abundant relative to time in both scenarios. In the Online Appendix, we extend this analysis to different compute levels. To distinguish between cases, we use the superscript ``$\star$'' for equilibrium outcomes with non-autonomous AI and ``$\post$'' for those with autonomous AI.

\subsection{The Impact of Non-Autonomous AI}

Since compute is abundant relative to time and AI is non-autonomous, the equilibrium rental rate of compute is zero, $r^{\star}=0$.  The reason is that some compute must remain idle. However, in contrast to the case of an autonomous AI, the wage of an individual with knowledge $\zbar$ is no longer equal to the equilibrium rental rate of compute ($w^{\star}(\zbar) \neq r^{\star}=0$). This difference occurs because an AI agent is no longer a perfect substitute for a human with knowledge $\zbar$.

For the following result, recall that $w(z)$ denotes the wage schedule of the pre-AI equilibrium:

\begin{prop} \label{prop:eqAInonautonomy} In the presence of a non-autonomous AI, there is a unique equilibrium. The equilibrium is efficient, maximizes labor income, and the rental rate of compute is $r^{\star}=0$. Moreover, \begin{itemize}[leftmargin=*,noitemsep]
\item If $\zbar \le w(0)$, then AI is not used in equilibrium, i.e., $W_a^{\star}=\emptyset$. Hence, human occupations and wages coincide with the pre-AI ones. 
\item If $\zbar>w(0)$, then only the least knowledgeable individuals use AI as a solver, i.e., $W_a^{\star}  \preceq (W_p^{\star} \cup I^{\star} \cup S_p^{\star})$, where all these sets are not empty except possibly $I^{\star}$. 
 \end{itemize} 
 In any case, \begin{enumerate}[leftmargin=*]  
 	\item Overall output is strictly higher with autonomous than non-autonomous AI.  \vspace{-2mm}
	\item Non-autonomous AI creates losers: \\ $\exists$ $z \in (0,1]$ such $w^{\star}(z) \le w(z)$ (with strict inequality if $\zbar>w(0)$). 
	\item The least knowledgeable benefit more from non-autonomous AI than from no AI or autonomous AI: \\ $\exists$ $\epsilon >0$ such that, for all $z \in [0,\epsilon)$, $w^{\star}(z) \ge \max\{ w(z),w^*(z)\}$ (with strict inequality if $\zbar>w(0)$).
	\item The most knowledgeable benefit more from autonomous AI than from non-autonomous AI: \\ $\exists$ $\epsilon >0$ such that, for all $z \in (1-\epsilon,1]$, $w^{\star}(z) \leq w^*(z)$ (with strict inequality if $z \neq 1$).
	\end{enumerate}
\end{prop}

Figure \ref{fig:na} illustrates Proposition \ref{prop:eqAInonautonomy}.\footnote{In the Online Appendix, we also examine how non-autonomous AI affects (i) human occupational choices, (ii) the distribution of firm size, productivity, and decentralization, and (iii) the productivity and span of control of workers and solvers who remain in their positions, compared to the pre-AI equilibrium.} According to this proposition, a non-autonomous AI is used in equilibrium only if it is sufficiently advanced (i.e., if $\zbar > w(0)$). In this case, individuals with knowledge below a certain threshold work in automated firms—using AI as a solver or co-pilot—while those above the threshold work in non-automated firms. Moreover, total output with non-autonomous AI is strictly lower than with autonomous AI because non-autonomy imposes a binding constraint on compute use, leaving some of it idle.

\begin{figure}[t!]
\begin{center}
  \begin{tikzpicture}[xscale=5.776095304, yscale=3.39770312]
 
\draw[->] (-.02+0, 0)--(1.03+0,  0);
\draw[->] (0+0, -.02)--(0+0,2.05-0);
\draw[dotted,thick] (0+0,0)--(1+0,1);
\node[right] at (1.03+0,0) {\Large $z$}; 
\draw[thick, gray, domain=0:{3-sqrt(5)}, samples=200] plot (\x+0, {3-sqrt(5)- (1/2*(3-sqrt(5))*(1+1/4*(3-sqrt(5))))/(1+1/2-1/2*(3-sqrt(5))) *(1-\x)  
+1/2*\x^2/2});
\draw[thick, gray, domain={3-sqrt(5)}:1, samples=200] plot (\x+0, {(1/2+2*(3-sqrt(5)))/(1+1/2*(1-(3-sqrt(5))))-sqrt(1+2*(3-sqrt(5))*2-2*\x*2)});
\draw[dashed] ({3-sqrt(5)+0},0)--({3-sqrt(5)+0},{3-sqrt(5)- (1/2*(3-sqrt(5))*(1+1/4*(3-sqrt(5))))/(1+1/2-1/2*(3-sqrt(5))) *(1-(3-sqrt(5)))  +1/2*(3-sqrt(5))^2/2}) ;
\draw (1+0,-0.02)--(1+0,0.02); 
\node[below] at (0+0,0) {$0$}; 
\node[below] at (1+0,0) {$1$}; 
\draw [decorate, decoration={brace, amplitude=10pt},gray] ({3-sqrt(5)+0},0)--(0+0,0);
\node[below] at ({(3-sqrt(5))/2+0},-.1) {$\color{gray} W$}; 
\draw [decorate, decoration={brace, amplitude=10pt},gray] (1+0,0) -- ({3-sqrt(5)+0},0);
\node[below] at ({3-sqrt(5)+(1-(3-sqrt(5)))/2+0.01+0},-.1) {$\color{gray} S$}; 
\node[above] at (.4+0,.625) {$w^{\star}(z)$}; 

\node[above,black] at (.80,1.5777) {$w^*(z)$}; 
\draw[ultra thick, domain=0:0.426, samples=100,black] plot (\x+0, {0.26656 + 0.26656*\x + 0.25000*\x^2});
\draw[ultra thick, black] (0.425+0,0.425)--(0.53311+0,0.53311);
\draw[ultra thick, domain=(0.53311-0.001):0.7009, samples=100,black] plot (\x+0, {1.5331 - sqrt(3.1323 - 3.9998*\x)});
\draw[ultra thick, domain=(0.70046-0.001):1, samples=100,black] plot (\x+0, {-1.4783 + 3.4783*\x});

\draw[ultra thick, dashed, domain=0:0.100765, samples=100] plot (\x+0, 0.425);
\draw[ultra thick, dashed, domain=0.100765:0.807142, samples=100] plot (\x+0, {-0.25000*\x^2 + 1.3746*\x - 0.11528 - 3.9069*(\x - 1.0006)*(\x^2 - 1.9994*\x + 0.99936)/sqrt(61.055*\x^2 - 122.11*\x + 61.055)});
\draw[ultra thick, dashed, domain=0.807142:1, samples=100] plot (\x+0, {1.7492 - sqrt(4.0372 - 4.0000*\x)});
  \node[left] at (0,0.425) {$\zbar$};


\draw[dotted] (1+0, 0)--(1+0,1.55);
\draw[dotted] (1+0, 1.68)--(1+0,2.1-0);
 \node[above] at (.425+0,0) {$\zbar$};
\draw (0.425+0,0.02)--(0.425+0,-0.02);
\draw[dashed] (0.425+0, .11)--(.425+0, .425)--(0+0, .425);
\node[right] at (.96,1.63) {$\color{gray} w(z)$}; 
\node[above] at ({1/2+0},-0.5) {  (a) Basic AI (i.e., $\zbar \in \mathrm{int} W$) };


\draw[->] (-.02+1.4, 0)--(1.03+1.4,  0);
\draw[->] (0+1.4, -.02)--(0+1.4,2.05-0);
\draw[dotted,thick] (0+1.4,0)--(1+1.4,1);
\node[right] at (1.03+1.4,0) {\Large $z$}; 
\draw[thick, gray, domain=0:{3-sqrt(5)}, samples=200] plot (\x+1.4, {3-sqrt(5)- (1/2*(3-sqrt(5))*(1+1/4*(3-sqrt(5))))/(1+1/2-1/2*(3-sqrt(5))) *(1-\x)  
+1/2*\x^2/2});
\draw[thick, gray, domain={3-sqrt(5)}:1, samples=200] plot (\x+1.4, {(1/2+2*(3-sqrt(5)))/(1+1/2*(1-(3-sqrt(5))))-sqrt(1+2*(3-sqrt(5))*2-2*\x*2)});
\draw[dashed] ({3-sqrt(5)+1.4},0)--({3-sqrt(5)+1.4},{3-sqrt(5)- (1/2*(3-sqrt(5))*(1+1/4*(3-sqrt(5))))/(1+1/2-1/2*(3-sqrt(5))) *(1-(3-sqrt(5)))  +1/2*(3-sqrt(5))^2/2}) ;
\draw (1+1.4,-0.02)--(1+1.4,0.02); 
\node[below] at (0+1.4,0) {$0$}; 
\node[below] at (1+1.4,0) {$1$}; 
\draw [decorate, decoration={brace, amplitude=10pt},gray] ({3-sqrt(5)+1.4},0)--(0+1.4,0);
\node[below] at ({(3-sqrt(5))/2+1.4},-.1) {$\color{gray} W$}; 
\draw [decorate, decoration={brace, amplitude=10pt},gray] (1+1.4,0) -- ({3-sqrt(5)+1.4},0);
\node[below] at ({3-sqrt(5)+(1-(3-sqrt(5)))/2+0.01+1.4},-.1) {$\color{gray} S$}; 

\draw[ultra thick, domain=0:.345025, samples=100] plot (\x+1.4, {.425+.425*\x});
\draw[ultra thick, domain=.345025:.85, samples=100] plot (\x+1.4, {.45476+.252488*\x+.25*\x^2});
\draw[ultra thick, domain=.85:.953, samples=100] plot (\x+1.4, {1.50498-sqrt(3.82899-4*\x)});
\draw[ultra thick, domain=.951623:1, samples=100] plot (\x+1.4, {-11.3333+13.3333*\x});
\node[above,black] at (1.4+.88,1.5777) {$ w^*(z)$};  

\node[above] at (.4+1.4,.86) {$w^{\star}(z)$}; 
\draw[ultra thick, dashed, domain=0:.73264, samples=100] plot (\x+1.4, 0.85);
\draw[ultra thick, dashed, domain=.73264:.982, samples=100] plot (\x+1.4, {.12191-.25*\x^2+1.1282*\x-(1.7376*(\x-.99917)*(\x^2-2.0008*\x+1.0008))/(sqrt(12.076*\x^2-24.153*\x+12.076)) }   );
\draw[ultra thick, dashed, domain=.982:1, samples=100] plot (\x+1.4, {1.2563+(\x-1.0001)/(sqrt(.25002-.25*\x))});

 
\node[right] at (1.4+.96,1.63) {$\color{gray} w(z)$}; 
\draw[dotted] (1+1.4, 1.68)--(1+1.4,2.1-0);
\draw[dotted] (1+1.4, 0)--(1+1.4,1.57);

\node[above] at (.85+1.4,0) {$\zbar$};
\draw (0.85+1.4,0.02)--(0.85+1.4,-0.02);
\draw[dashed] (.85+1.4, .11)--(.85+1.4, .85)--(0+1.4, .85);
\node[above] at ({1/2+1.4},-0.5) {  (b) Advanced AI (i.e., $\zbar \in \mathrm{int} S$) };  
  \node[left] at (1.4,0.85) {$\zbar$};
 
  \end{tikzpicture}
  \captionsetup{justification=centering}
  \caption{Non-Autonomous AI vs. Autonomous AI vs. No AI \\ \justifying 
  \vspace{0.5mm}
 \footnotesize \noindent  \textit{Notes}. Distribution of knowledge: $G(z)=z$. Parameter values: $h=1/2$. Moreover, for panel (a), $\zbar=0.425$, while for panel (b), $\zbar=0.85$. The thick gray line represents the pre-AI equilibrium wage function \( w \). The thick black dashed line depicts the equilibrium wage function with non-autonomous AI, \( w^{\star} \). The thick black line represents the equilibrium wage function with autonomous AI, \( w^{\post} \).} \label{fig:na}
 \end{center}
\end{figure}

In terms of wages, the introduction of non-autonomous AI creates both winners and losers, similar to autonomous AI. However, the distribution of gains and losses changes dramatically. Non-autonomous AI always benefits the least knowledgeable individuals relative to the pre-AI equilibrium. Moreover, these benefits exceed those of autonomous AI. In contrast, non-autonomous AI may lead to gains or losses compared to the pre-AI scenario for the most knowledgeable individuals. In any case, their wages with non-autonomous AI are always lower than with autonomous AI.

Intuitively, non-autonomous AI benefits the least knowledgeable individuals the most because it functions solely as a solver. This eliminates competition between these individuals and AI for production work and enables them to solve complex problems without human assistance. Moreover, non-autonomous AI’s lower opportunity cost—stemming from its inability to produce independently—enhances this advantage by allowing the least knowledgeable to capture a larger share of AI-generated benefits. For the most knowledgeable individuals, the opposite occurs because non-autonomous AI cannot handle routine work and may compete with them in advisory roles.

As noted earlier, we limit firms to a maximum of two layers to ensure a meaningful comparison of equilibrium outcomes between autonomous and non-autonomous AI. This restriction provides an additional benefit: it implies that problems within the organization can be escalated only once, effectively introducing a cost of seeking help beyond the helping cost \(h\) (which is borne solely by the recipient of the question). Notably, this implies that even with free compute, individuals with slightly less knowledge than AI prefer to seek help from another human rather than from AI (see Figure \ref{fig:na}), as the probability that AI provides them with helpful assistance is low.\footnote{The equilibrium is continuous with respect to the availability of compute, so all our results hold when the price of compute is strictly positive but sufficiently close to zero. In the Online Appendix, we also examine the case where compute is significantly less abundant. The results of this section extend to that scenario as well, albeit with certain qualifications.}

We believe this is a reasonable outcome, as a similar situation would arise if problems could be escalated multiple times but the agents seeking assistance bore part of the time cost associated with requesting help. Without these constraints—such as with an arbitrary and endogenous number of layers and no time cost for the asking party—all individuals less knowledgeable than AI would always rely on the system for help. In this scenario, introducing AI effectively shifts the knowledge of all agents with \(z < \zbar \) towards \(\zbar\). Nevertheless, the key result of this section—that the least knowledgeable individuals prefer non-autonomous AI while the most knowledgeable favor autonomous AI—remains valid even under this alternative case.\footnote{With more than two layers, an advanced non-autonomous AI could further amplify the expertise of the most knowledgeable individuals by giving them access to workers that require less help (as workers would first seek AI assistance before turning to them). However, the key insight remains: the most knowledgeable individuals still prefer autonomous AI over non-autonomous AI, as the former can also perform routine work for them.}
 
\section{Discussion and Final Remarks} \label{sec:conclusion}

To conclude, we discuss additional implications of our analysis and link our findings to emerging evidence and stylized facts about AI's effects on the labor market. 

\vspace{2mm} 

\noindent \textit{Co-pilots vs. Co-workers}.---Our analysis uncovers distinct labor market implications when comparing AI co-pilots---which provide problem-solving assistance---to AI co-workers---which independently execute production work. Broadly speaking, AI co-pilots primarily benefit less knowledgeable individuals, reducing inequality in performance and earnings. In contrast, AI co-workers offer the greatest advantages to highly knowledgeable individuals, amplifying the value of their expertise. Co-pilots emerge in equilibrium when AI is either (i) not autonomous or (ii) autonomous but sufficiently advanced, whereas co-workers can arise for any knowledge level but only in the case in which AI is autonomous.

These findings are particularly relevant given the emerging evidence on AI’s labor market effects. Several studies suggest that AI disproportionately benefits individuals with the least knowledge and reduces performance inequality \citep[e.g.,][]{dell2023navigating,noy2023experimental,peng2023impact,wiles2023algorithmic,demming,brynjolfsson2023generative}.\footnote{For instance, \cite{brynjolfsson2023generative} reports that deploying an AI-based conversational assistant significantly improved productivity for less-skilled agents in a Fortune 500 customer support center, with minimal impact on experienced ones. Similarly, \cite{demming} reported that AI assistance was more beneficial for lower-ability individuals than for higher-ability ones when tasked with estimating whether people in photographs appeared to be over 21 years old.} However, these studies focus on AI co-pilots. While our framework supports these conclusions, it also highlights that they may not extend to autonomous AI that can perform independent knowledge work, such as conducting research, drafting documents, or performing data analysis. 

Moreover, non-experimental evidence suggests that firms perceive current iterations of generative AI more as a basic co-worker than as a co-pilot. For instance, using high-frequency job posting data from LinkUp, \cite{berger2} document a marked decline in job postings for white-collar roles requiring low levels of knowledge, such as entry-level office jobs, following the introduction of ChatGPT in November 2022. At the same time, job postings for high-knowledge roles, such as executive positions, increased. This trend indicates that firms view generative AI as a technology that complements high-skilled knowledge workers while substituting for their low-skilled counterparts. 

Further empirical studies that disentangle and estimate the distinct labor market effects of AI co-workers and AI co-pilots represent an important avenue for future research.

\vspace{2mm}

\noindent \textit{AI vs.Traditional Robots and Physical AI}.---Automation technologies that replace manual labor, such as traditional robots and physical AI, differ fundamentally from those that automate cognitive work, such as AI and enterprise software. This difference arises because physical machines must be tailored to their specific operating environments, making them best described as a “putty-clay” technology. In contrast, digital machines—such as AI agents—exhibit “putty-putty” characteristics because they rely on compute, which is general-purpose and can be rented on demand. This allows software improvements to be immediately deployed across all available hardware.\footnote{Putty-clay models assume that capital and technology are flexible only prior to investment decisions, whereas putty-putty models allow for flexibility both before and after investments are made \citep{atkeson1999models,martinez2021putty}.}

A key implication of this distinction is that traditional robots and physical AI are likely to exist in vintages with varying capabilities, while compute is allocated to the most advanced and highest-performing digital machines.

\vspace{2mm}

\noindent \textit{AI vs.Offshoring and Immigration}.---The shock induced by AI is different from classical labor shocks such as offshoring or immigration. Non-autonomous AI, for instance, has no clear analog in these frameworks. Moreover, even for autonomous AI, the introduction of the technology diverges significantly from offshoring or immigration. The key distinction lies in AI's scalability. Once knowledge is encoded into an AI system, it can be deployed across all units of compute, which are widely available. This scalability amounts to the introduction of a large population of homogeneous agents. 

In contrast, large human populations are inherently heterogeneous in terms of tacit knowledge---different individuals undoubtedly have different experiences. Consequently, offshoring or immigration can involve the introduction of a relatively small population of homogeneous agents or the introduction of a large population of heterogeneous agents, but not the introduction of a large population of homogeneous agents. As we formally show in the Online Appendix, this difference leads to qualitatively different outcomes.

\vspace{2mm} 

\noindent \textit{AI vs. ICTs}.--- A significant implication of our model is that not all firms adopt AI to \textit{automate knowledge work}. This characteristic makes AI, as an automation technology, fundamentally different from earlier waves of digitalization—such as the introduction of computers or email—which saw near-universal adoption across firms. Partial adoption is a common characteristic of automation technologies. Both empirical observations and standard models of automation \citep[e.g.,][]{acemoglu2022automation} suggest that it is often not cost-effective to automate all tasks that can potentially be automated.

For autonomous AI, partial adoption occurs because deploying AI involves a strictly positive opportunity cost: the foregone output that an autonomous AI agent could generate as an independent producer. As a result, adopting autonomous AI may not be economically viable for all firms, consistent with the evidence presented by \citet{svanberg2024beyond}. For non-autonomous AI, partial adoption arises even when the opportunity cost is zero. This is because AI’s advice is useless in firms with very knowledgeable human workers.

However, the value of AI extends beyond automating knowledge work. Thus, firms may adopt AI technologies for purposes unrelated to automation, such as reducing communication costs or lowering knowledge acquisition costs. The effects of such technologies have been extensively studied \citep[see, e.g.,][]{garicano2006organization,antras2008organizing,bloom2014distinct,garicano2015knowledge}, and are fundamentally different from the ones of AI as an automation technology.\footnote{For instance, better communication technologies---modeled as a reduction in the communication cost $h$---always displace humans from specialized problem-solving towards routine work as the very best solvers can now supervise larger teams. In contrast, the displacement generated by AI as an automation technology depends on its autonomy and its problem-solving capabilities.}

\vspace{2mm} 

\noindent \textit{AI in Advanced vs. Developing Economies}.--- An important implication of our analysis is that AI's effects depend on its capabilities relative to human pre-AI occupations. Since these occupations are endogenous—shaped by the knowledge of the population and the level of communication technologies—the same AI technology can lead to different outcomes in advanced and developing economies.

Indeed, in economies with superior communication technologies or a more knowledgeable population, the knowledge threshold required to become a solver in the pre-AI equilibrium is higher (see the Online Appendix for details). As a result, the same autonomous AI technology might be considered ``basic'' in an advanced economy—displacing humans from routine knowledge work to complex problem-solving—while being regarded as ``advanced'' in a developing economy, thereby displacing humans in the opposite direction.

\vspace{2mm} 

\noindent \textit{Autonomy vs. No Autonomy: Policy Implications}.--- The rise of AI and its potential to transform labor markets has captured policymakers' attention, as evidenced by the recent report of \citet{CEA} and the \citet{EU}. Our findings reveal key tradeoffs in regulating AI autonomy. While autonomous AI generates higher overall output than non-autonomous AI, it also exacerbates labor income inequality. 

Using this framework to quantify the tradeoff between output and inequality associated with autonomous and non-autonomous AI is an important direction for future research.

\newpage 
{\small
\bibliographystyle{ecta}
\bibliography{AI}}

\end{document}